\documentclass[12pt,rmp,tightenlines]{revtex4}
\usepackage[final]{graphicx}  
\usepackage{amsmath,amssymb} 
\usepackage{indentfirst}
\usepackage{epsf}
\usepackage{psfrag}
\usepackage{float}

\begin{document}

\title{\vskip1.55cm Brane Universe: Global Geometry}

\author{Victor Berezin}
\affiliation {Institute for Nuclear Research, Russian Academy
of Sciences,\\
60th October Anniversary Prospect, 7a, 117312, Moscow, Russia\\
e-mail: berezin@ms2.inr.ac.ru}

\begin{abstract}
{\it The global geometries of bulk vacuum space-times in the
brane-universe models are investigated and classified in terms of
geometrical invariants. The corresponding Carter-Penrose diagrams
and embedding diagrams are constructed. It is shown that for a given
energy-momentum induced on the brane there can be different types of
global geometries depending on the signs of a bulk cosmological term
and surface energy density of the brane (the sign of the latter does
not influence the internal cosmological evolution). It is shown that
in the Randall-Sundrum  scenario it is possible to have an
asymmetric hierarchy splitting even with a $Z_2$-symmetric matching
of "our" brane to the bulk.}
\end{abstract}
\maketitle


In this talk we would like to investigate the possible global
geometries of the so-called "brane universe" scenarios \cite{c1,c2,c3}, in which our
Universe is supposed to be a thin shell, "membrane", embedded into
the space-time of larger number of dimensions, "bulk".

Our strategy is to simplify everything as much as possible and
construct some exactly solvable model, because only the thorough
investigation of such models is the source of the physical
intuition. So, let us consider a $(N+1)$-dimensional space-time
containing a $N$-dimensional brane (thin shell) with the metric
\begin{equation}
\label{m}
ds^2 = g_{\mu \nu}(y) dy^{\mu} dy^{\nu}.
\end{equation}
Having in mind the very name of the Conference, we demand the brane
to be time-like and have the so-called cosmological symmetry, i.e.,
homogeneity and isotropy. Moreover, we assume (and this is the first
step in our simplification process) that outside the brane the
"bulk" geometry possesses the same symmetry, in other words,
locally, the bulk geometry does not depend on the place of the
brane. This means, that throughout the whole $(N+1)$-dimensional
manifold we can introduce the normal Gaussian coordinate system in
which the metric (\ref{m}) takes the form
\begin{eqnarray}
\label{gm}
ds^2 &=& - dn^2 + \gamma_{ij}(n,x) dx^i dx^j \nonumber \\
&=& - dn^2 + B^2(n,t) dt^2 - A^2(n,t) dl^2_{N-1},
\end{eqnarray}
where $i,j$ take the values $(0,2,...N)$, and $dl^2_{N-1}$ is the
Robertson-Walker unit line element of a homogeneous space,
\begin{equation}
\label{RW}
dl^2_{N-1} = \frac{dr^2}{1 - k r^2} + r^2 d\Omega^2_{N-2}
\end{equation}
with $d\Omega^2_{N-2}$ representing the line element of the unit
$(N-2)$-dimensional sphere. For $k=+1$ the homogeneous space is the
unit $(N-1)$-dimensional sphere, for $k=-1$ it is the rotational
hyperboloid, and $k=0$ means that such a space is flat. For a while
we suppose that there exists only one brane in a bulk and put it at
zero value of the normal coordinate $n=0$. Then, the general form of
the energy momentum tensor $T_{\mu \nu}$ is
\begin{eqnarray}
\label{emt}
T_{\mu \nu} = S_{\mu \nu} \delta(n) &+& \left [ T_{\mu \nu} \right] \Theta (n)
+ T_{- \mu \nu} , \nonumber\\
\left[ T_{\mu \nu} \right] = T_{+ \mu \nu} - T_{- \mu \nu}.
\end{eqnarray}
Here $S_{\mu \nu}$ is the surface energy-momentum tensor on the
brane, square brackets $[\:]$ denote a jump of some quantity across
the shell (brane), $[X] = (X_+ - X_-)$, indices $"\pm"$ indicate the
$n>0 ("+")$ and the $n<0 ("-")$ regions outside the shell, $\delta
(n)$ and $\Theta (n)$ are conventional Dirac's and step functions.
If $S_{\mu \nu} \ne 0$, the hyper-surface $n = const$ is singular,
otherwise it is regular. Introducing the extrinsic curvature tensor
$K_{ij} = - \frac{1}{2} \frac{\partial \gamma_{ij}}{\partial n} = -
\frac{1}{2} \gamma_{ij,n}$ for the $N$-dimensional hyper-surfaces $n
= const$ (both singular and regular ones), we are able to separate
the Einstein equations into three groups according to the above
decomposition of the energy-momentum tensor (everywhere $G$ is the
non-renormalized $(N+1)$-dimensional gravitational constant).

1. $\left( ^n_n \right)$ -equations:
\begin{eqnarray}
\label{delta}
S^l_j \left\{ K^j_l \right\} &+& \left[ T^n_n \right] = 0 \,,\nonumber \\
\frac{1}{2} K^l_{-j} K^j_{-l} &-& \frac{1}{2} K^2_{-}
- \frac{1}{2}\,^{(N)}R = 8 \pi \, G \,T^n_{-n}
\end{eqnarray}
and analogous equations in $"+"$-region, $K = K^j_j$ is the trace of
the extrinsic curvature tensor. The parenthesis means $\{X\} =
\frac{1}{2} (X_+ + X_-)$, and $ ^{(N)}R$ is the $N$-dimensional
Ricci scalar on every hyper-surface $n = const$.

2. $\left( ^n_i \right)$-equations:
\begin{eqnarray}
\label{theta}
S^l_{i|l} &+& \left[ T^n_i \right] = 0 \,, \nonumber \\
K^l_{-i|l} &-& K_{-|i} = 8 \pi \, G \,T^n_i \,,
\end{eqnarray}
where the vertical line denotes a covariant derivative with respect
to the $N$-dimensional metric $\gamma_{ij}(x,n)$, and for the sake
of brevity we will not mention anymore the equations in
$"+"$-region. The first of this set of equations is nothing more but
the continuity equation for $S^j_i$.

3. $(i k)$-equations:
\begin{eqnarray}
\label{cont}
&-& \left( \left[ K_{ik} \right] - \gamma_{ik} \left[ K \right] \right) =
8 \pi \,G\, S_{ik} \,, \nonumber \\
&2& S^l_i \left\{ K_{lk}\right\} + 2 S^l_k \left\{ K_{il}\right\} -
\frac{3}{N-1} S \left\{ K_{ik}\right\} - S_{ik} \left\{ K \right\} +
\gamma_{ik} S^l_j \left\{ K^j_l\right\} -
\frac{1}{N-1} \gamma_{ik} S \left\{ K \right\} = \left[ T_{ik}\right] \,,
\nonumber \\
&-& \left( K_{-ik,n} - \gamma_{ik} K_{-,n} + 2 K_{-il} K^l_{-k} -
K_{-ik} K + \frac{1}{2} \gamma_{ik} \left( K^l_{-j} K^j_{-l} + K\right) \right) +
 ^{(N)}G_{ik} = 8 \pi \,G \, T_{-ik} \,.
\end{eqnarray}
The equations in the first line are known as the Israel's equations.
$ ^{(N)}G_{ik}$ is the $N$-dimensional Einstein tensor on every
hyper-surface $n = const$. The last equation can also be written in
the form
\begin{equation}
\label{ind}
 ^{(N)}G_{ik} = 8 \pi \,G \,T_{-ik} + T^{ind}_{-ik} \,.
\end{equation}
Using the proclaimed cosmological symmetry it is easy to calculate
both $ ^{(N)}R$ and $ ^{(N)}G_{ik}$. For each hyper-surface $n =
const$ we introduce the cosmological time by the relation $d\tau_n =
B(n,t) dt$ and the scale factor $a(\tau_n) = A(n,t)$, then, by
symmetry, $ ^{(N)}G^2_2 = ^{(N)}G^3_3 = ... = ^{(N)}G^N_N$, and
\begin{eqnarray}
\label{RG}
 ^{(N)}R &=& - \left( N - 1\right) \left( \frac{2 a_{\tau \tau}}{a} +
 \left( N - 2\right) \frac{a^2_{\tau} + k}{a^2}\right) \,, \nonumber \\
  ^{(N)}G^0_0 &=& \frac{(N-1) (N-2)}{2} \frac{a^2_{\tau} + k}{a^2} \,,
  \nonumber \\
  ^{(N)}G^2_2 &=& \frac{N - 2}{2} \left( \frac{2 a_{\tau \tau}}{a} +
 (N -3) \frac{a^2_{\tau} + k}{a^2}\right)\,
\end{eqnarray}
where $a_{\tau} = \frac{da}{d\tau}, \; a_{\tau \tau} = \frac{d^2
a}{d\tau^2}$

The cosmological principle allows us to use also yet another
technique in investigation of the global geometry. This is the
so-called $(d+2)$-decomposition, and it is not related to the
singular brane, but deals exclusively with the invariants of the
bulk geometry. So, let us start. The metric of any
$(d+2)$-dimensional space-time which is a direct product of a
$d$-dimensional homogeneous space (cosmological symmetry!) and a
two-dimensional space-time can be written in the form
\begin{equation}
\label{d2}
ds^2 = \gamma_{AB}(x) dx^A dx^B - R^2 (x) dl^2_d \,,
\end{equation}
where $dl^2_d$ - the unit Robertson-Walker line element for a
homogeneous space with the curvature $d(d-1)k, \; k = \pm 1,\,0, \;
A = 0,1$ ($"0"$ for some time coordinate $t$, $"1"$ for some radial
coordinate $q$), $\gamma_{AB}$ is a two-dimensional metric tensor,
and $R(x)$ is the radius or, in other words, scale factor, of the
$d$-dimensional homogeneous space. Due to the general covariance a
two-dimensional geometry is locally determined actually by only one
function of two variables $t$ and $q$. For the $(d+2)$-dimensional
manifold with cosmological symmetry we need, therefore, to know only
two functions of two variables. Naturally, one of them is the radius
$R(t,q)$ which is invariant under $(t,q)$-transformations. Surely,
we want the second function to be also an invariant. Geometrically,
the best choice is the squared normal to the surfaces $R = const$.
So, we define our second function as
\begin{equation}
\label{Delta}\Delta = \gamma^{AB} R_{,A} R_{,B} \,
\end{equation}
where comma means a partial derivative, $R_{,a} = \frac{\partial
R}{\partial x^A}$. Remarkably enough that, using these two
invariants, we can rewrite the two-dimensional part of the Einstein
equations in this case in a very convenient vector-like form
\begin{equation}
\label{vec}
\left( R^{d-1} \left( \Delta + k \right)\right)_{,A} =
\frac{16 \pi \, G}{d} R^d \left( T \, R_{,A} - T^B_A \, R_{,B}\right)\,,
\end{equation}
where $T = T^C_C$. The third equation, for $A \ne B$
\begin{equation}
\label{AnB}
\gamma^{AC} \, R_{||CB} = - \frac{8 \pi \, G}{d}\, R\, T^A_B \,,\;\;\; A \ne B\,,
\end{equation}
can also be obtained as an integrability condition for the above
vector equation. The double vertical line here denotes a covariant
derivative with respect to the two-dimensional metric $\gamma_{AB}$.
Our invariant $\Delta$ brings a very important geometrical
information. Note, first of all, that for the flat Minkowskian
space-time (of any dimension) $\Delta = - 1$. But in the curved
space-time $\Delta$ is no more a constant and can be both negative
and positive. If it is negative, we can choose the radius $R$ as a
spatial coordinate $(\dot R = 0, \, (R^{\prime})^2 = 1
\Longrightarrow \Delta = \gamma^{11} (R^{\prime})^2 = \gamma^{11} <
0)$ like in the flat space-time, and the surfaces $R = const$ are
time-like. Such regions are called the $R$-regions \cite{c4}. Moreover, in
these regions $R_{,q}$ cannot change its sign, therefore we may have
either $R_{,q} > 0$ in the $R_+$-regions, or $R_{,q} < 0$ in the
$R_-$-regions. Analogously, if $\Delta > 0$, the surfaces $R =
const$ are space-like, and the radius $R$ can be used as a time
coordinate, these regions are called the $T$-regions. And, again,
now the sign of $R_{,t}$ cannot be changed, so, there are
$T_+$-regions with $R_{,t > 0}$ (inevitable expansion) and
$T_-$-regions with $R_{,t < 0}$ (inevitable contraction). The $R$-
and $T$-regions are separated by hyper-surfaces $\Delta = 0$ called
the apparent horizons. The global geometry of the space-time
manifolds with cosmological symmetry is, therefore, the set of
$R_{\pm}$- and $T_{\pm}$-regions separated by the apparent horizons
$\Delta = 0$ \cite{c5,c6}. Of course, not all such sets are physical. As a
selection rule we will use the physical principle of geodesic
completeness: any null or time-like geodesics must start and end
either at infinities or at singularities where the Riemann curvature
tensor becomes divergent.

It is time to make the next (second) step in simplification of our
model. We assume that outside the brane at $n = 0$ the space-time is
a vacuum with cosmological constant $\Lambda$, thus, the
energy-momentum tensor for $n \ne 0$ has the form of an invariant
tensor
\begin{equation}
\label{lambda}
8 \pi \, G T^{\nu}_{\mu} = \Lambda \, \delta^{\nu}_{\mu} \,,
\end{equation}
in particular, $8 \pi \, G T^B_A = \Lambda \delta^B_A, \; 8 \pi \, G
\,T^C_C = 2 \Lambda $. The vector-like equations are easily
integrated now to give
\begin{equation}
\label{deltam}
\Delta = -k + \frac{2\,G\, m}{R^{N-2}} + \frac{2}{N (N-1)} \Lambda \, R^2 ,
\end{equation}
here $m$ is an integration constant with the dimension of mass. Note
that here the invariant $\Delta$ is actually a function of one
variable - invariant radius $R$. In such a case it is easy to write
explicitly the two-dimensional line element separately in $R$- and
$T$-regions. If $\Delta < 0$ ($R$-region), we can choose the radius
as the spatial coordinate $q$ (or $-q$). Then,
\begin{equation}
\label{bulkR}
ds^2_2 = \left( -\Delta \right) dt^2 - \frac{d R^2}{\left( - \Delta\right)}\,.
\end{equation}
In what follows we will need yet another form of the line element,
namely, the conformally flat one. For this let us introduce the
function $R^{\star}(R)$ by the relation
\begin{equation}
\label{rstar}
dR^{\star} = \pm \frac{dR}{\left| \Delta \right|} \,,
\end{equation}
then
\begin{equation}
\label{cflat}
ds^2_2 = \left( - \Delta\right) \left( dt^2 - d{R^{\star}}^2 \right) \,.
\end{equation}
In $T$-regions $\Delta > 0$ and we can choose $R$ (or $-R$) as a
time coordinate, for the line element one gets
\begin{equation}
\label{cflatt}
ds^2_2 = \frac{dR^2}{\Delta} - \Delta dq^2 = \Delta \left( d {R^{\star}}^2 -
dq^2\right) \,,
\end{equation}
here $R^{\star} \, ( - R^{\star})$ is a time coordinate, and $q$ a
spatial coordinate of the Minkowskian (flat) two-dimensional
space-time. We assume that there can be only one singular shell in
the whole $N+1$-dimensional space-time, namely, our brane. Then, to
avoid a singularity at $R = 0$ we have to put $m = 0$. In this case
the value $k = \pm 1, 0$ should be the same everywhere, it is a
global property. For different $k$ we obtain completely different
bulk space-times. We see, that such a global feature dictates the
spatial curvature on the brane. The latter can be determined, in
principle, by making measurements on the brane itself. And this is
the way to know something about the bulk geometry (if, of course, we
have some other evidences that the brane universe hypothesis is
true). The case of several branes will be briefly discussed later.

Let us go further and make use of the Eqn(\ref{AnB}). Substituting
in it the two-dimensional part of the metric (\ref{gm}), namely,
$ds^2_2 = - dn^2 + B^2 (n,t) dt^2$, we obtain
\begin {equation}
\label{ft} \Delta = \frac{1}{B^2 (n,t)} R^2 (n,t)_{,t} - R^2
(n,t)_{,n} = f^2(t) - R^2_{,n} \,,
\end{equation}
where $f(t)$ is some function of time coordinate only. From this we
have
\begin{equation}
\label{Rn}
R_{,n} = \pm \sqrt{f^2 (t) - \Delta} = \sigma \sqrt{f^2 (t) - \Delta}\,.
\end{equation}
We introduced new and very important sign function $\sigma$. It
shows whether radii increase with $n (\sigma = +1)$, or they
decrease $(\sigma = -1)$. It is clear from the definition that in
$R_+$-regions $\sigma = +1$, and $\sigma = -1$ in $R_-$-regions. In
$T$-regions $\sigma$ may change the sign. Thus, this sign will point
at the region where exactly the brane is matched to the bulk. This
last equation together with the fact that the invariant $\Delta$,
Eqn.(\ref{deltam}), depends only on the radius $R$ allows us to
obtain the solution $R(n,t)$ as an explicit function of the normal
coordinate $n$. To have the full information we need the equations
on the brane at $n = 0$. Remembering that $R(n,t) = A(n,t)$ and
$f(t) = \frac{R_{,t}}{B} = a_{\tau}$, we are able to calculate the
extrinsic curvature tensor $K_{ij}$ and the induced energy-momentum
tensor on the brane $T^{ind}_{-ij}$. We need also a relation between
the coordinate time $t$ and the cosmological time $\tau$ on the
brane. Using the freedom (gauge) in defining the coordinate time, we
can always put $B(0,t) = 1$, in other words, $t = \tau$ for $n = 0$
on the brane. Let us remind that the Israel's equations (matching
conditions) give some relations between the jump in extrinsic
curvature tensor and the surface energy-momentum tensor $S^j_i$. It
is easy to show that this tensor determines also (together with the
bulk cosmological constant) the induced energy-momentum tensor. We
see now that, given the surface energy-momentum tensor $S^j_i$, the
sign of the spatial curvature $k$ and the value and the sign of the
cosmological constant $\Lambda$ we can construct both the global
geometry of the bulk and the trajectory of the brane. Therefore, we
will know the complete geometry of the whole space-time.

Of course, in general, it is still impossible to get the solution in
a closed form. Hence, we need to simplify the model further. And as
the final step, we restrict ourselves to investigation of the vacuum
shells. Namely, we choose the following equation of state
\begin{equation}
\label{eos}
S^0_0 = S^2_2 \,.
\end{equation}
From the first of Eqns.(\ref{theta}) we have $S^0_0 = const$, and
the set of equations we need, looks as follows
\begin{eqnarray}
\label{set}
R_{,n} ({\pm}) &=& \sigma_{\pm} \sqrt{f^2 (t) + k -
\frac{2 \Lambda}{N(N-1)} R^2} \,, \nonumber \\
- \left[ \frac{R_{,n}}{R}\right] &=& \frac{1}{R} \left( R_{,n}(-) -
R_{,n}(+)\right) = \frac{8 \pi \, G}{N-1}\, S^0_0 \,, \nonumber \\
S^0_0 &=& const \,, \;\; R (0,t) = a (\tau)\,, \;\; \tau = t \,, \nonumber \\
\frac{(N-1)(N-2)}{2} \frac{a^2_{\tau} + k}{a^2} &=&
\frac{N-2}{N} \left( \Lambda + \frac{N}{2 (N-1)}
\left( 4 \pi \, G\right)^2 \left( S^0_0\right)^2 \right) \,.
\end{eqnarray}
Note, first of all, that the values of $R_{,n}$ on different sides
of the brane differ by their sign only, this is the consequence of
our assumption to have $m = 0$ everywhere in the bulk. Therefore,
$\sigma_- = - \sigma_+$, and we automatically obtain the
$Z_2$-symmetric brane. Moreover, the signs of $S^0_0$ and $\sigma_-$
are the same, the latter affects the matching of the brane to the
bulk, but not the evolution inside the shell. Let us now solve the
set of equations (\ref{set}), considering all the possibilities one
by one.

We begin with positive cosmological constant, $\Lambda > 0$.
Introducing (for brevity) the so-called cosmological radius $R_0 =
\sqrt{\frac{N(N-1)}{2 \Lambda}}$ and suppressing $(\pm)$ indices we
get from the first of Eqns.(\ref{set})
\begin{equation}
\label{R}
R = R_0 \, \sqrt{f^2(t) + k} \sin{\left(\frac{\sigma n}{R_0} +
\varphi (t) \right)} \,,
\end{equation}
where $\varphi (t)$ is an another function of time. On the brane at
$n = 0$ the following equations are valid $(\sigma = \sigma_- = -
\sigma_+)$:
\begin{eqnarray}
\label{setb}
\frac{\sigma}{R_0} \cot{\varphi} &=& \frac{4 \pi \, G}{N-1} S^0_0 \,,
\nonumber \\ \frac{a^2_{\tau} + k}{a^2} &=& \frac{2 \Lambda}{N(N-1)} +
\left(\frac{4 \pi \, G}{N-1} \right)^2 \left( S^0_0 \right)^2 =
\frac{1}{R_0^2 \sin^2 {\varphi}} \,.
\end{eqnarray}
Since $S^0_0 = const$, then $\varphi = \varphi_0 = const$. We see
also that the value of $\sigma (= \sigma_- = - \sigma_+)$ depends on
the sign of $S^0_0$ and affects crucially the matching of our brane
to the bulk. For different values of $k = \pm 1, 0$ only the time
dependent pre-factor is changed:
\begin{eqnarray}
\label{cases}
R &=& R_0 \sin{\left( \frac{\sigma n}{R_0} +
\varphi_0\right)} \begin{cases} \cosh{\frac{t}{a_0}} \,,
&\text {for \; \;$k = +1$} \nonumber \\
e^{\frac{t}{a_0}} \,, &\text {for \; \;$k = 0$} \nonumber \\
\left| \sinh {\frac{t}{a_0}}\right| \,, &\text {for \; \;$k = -1$} \nonumber
\end{cases} \nonumber \\
a_0 &=& R_0 \sin{\varphi_0}
\end{eqnarray}
It is useful to visualize the matching by plotting the above
function $R = R(n)$ for some moment of time $t = const$. We have for
$\sigma = \sigma_- = +1 (\sigma_+ = -1)$ (Figs.1 and 2):
\begin{figure}[H]
\unitlength=1in
\begin{center}
\psfragscanon
\psfrag{y}{$\frac{R}{R_0}$}
\psfrag{x}{$\frac{n}{R_0}$}
\psfrag{z}{$0$}
\psfrag{phi0}{$-\phi_0$}
\psfrag{phi1}{$\pi-\phi_0$}
\psfrag{sigma}{$\sigma=+1$}
\psfrag{n}{$n<0$}
\includegraphics[width=3.5in]{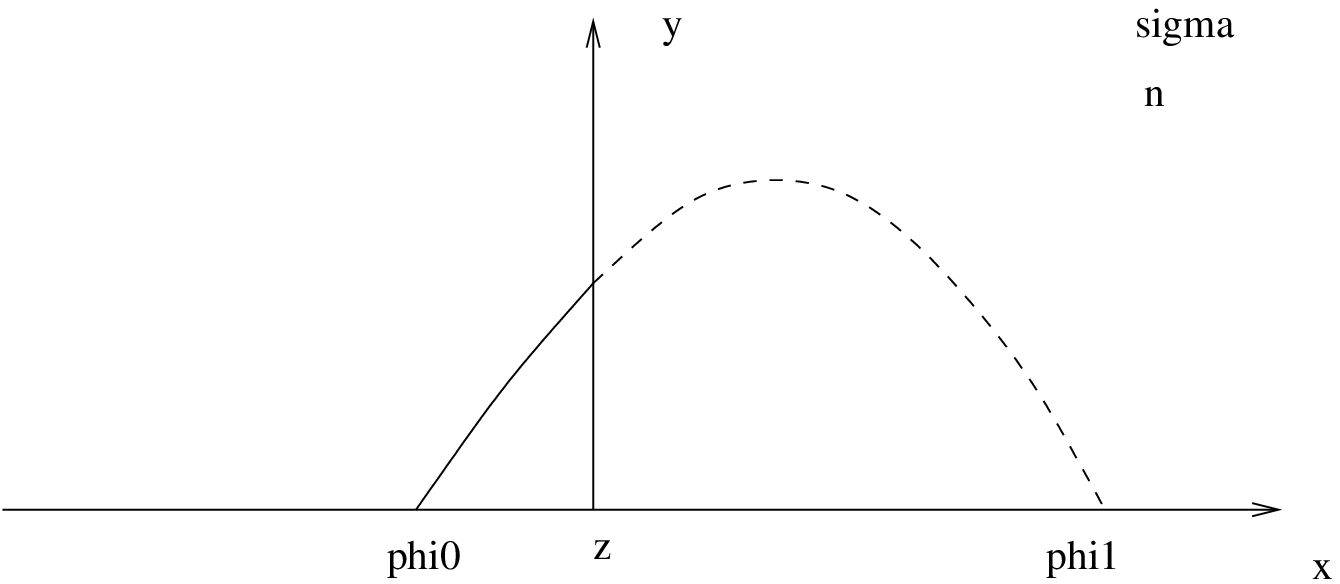}
\end{center}
\caption{}
\end{figure}
\begin{figure}[H]
\unitlength=1in
\begin{center}
\psfragscanon
\psfrag{y}{$\frac{R}{R_0}$}
\psfrag{x}{$\frac{n}{R_0}$}
\psfrag{z}{$0$}
\psfrag{phi0}{$-\pi+\phi_0$}
\psfrag{phi1}{$\phi_0$}
\psfrag{sigma}{$\sigma=+1$}
\psfrag{n}{$n>0$}
\includegraphics[width=3.5in]{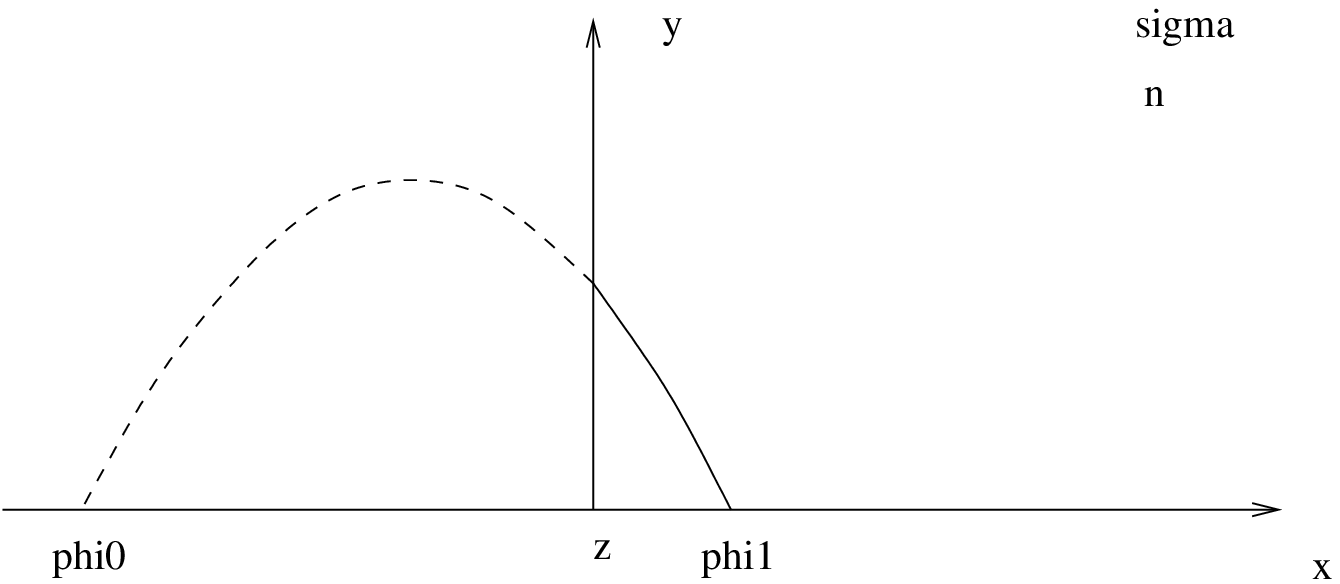}
\end{center}
\caption{}
\end{figure}
The dashed curves show a continuation of the function $R(n)$ beyond
the shell. Combining these two Figures we get Fig.3:
\begin{figure}[H]
\unitlength=1in
\begin{center}
\psfragscanon
\psfrag{y}{$\frac{R}{R_0}$}
\psfrag{x}{$\frac{n}{R_0}$}
\psfrag{z}{$0$}
\psfrag{phi0}{$-\phi_0$}
\psfrag{phi1}{$\phi_0$}
\psfrag{sigma}{$\sigma=+1$}
\psfrag{sh}{$\mbox{shell}$}
\includegraphics[width=3.5in]{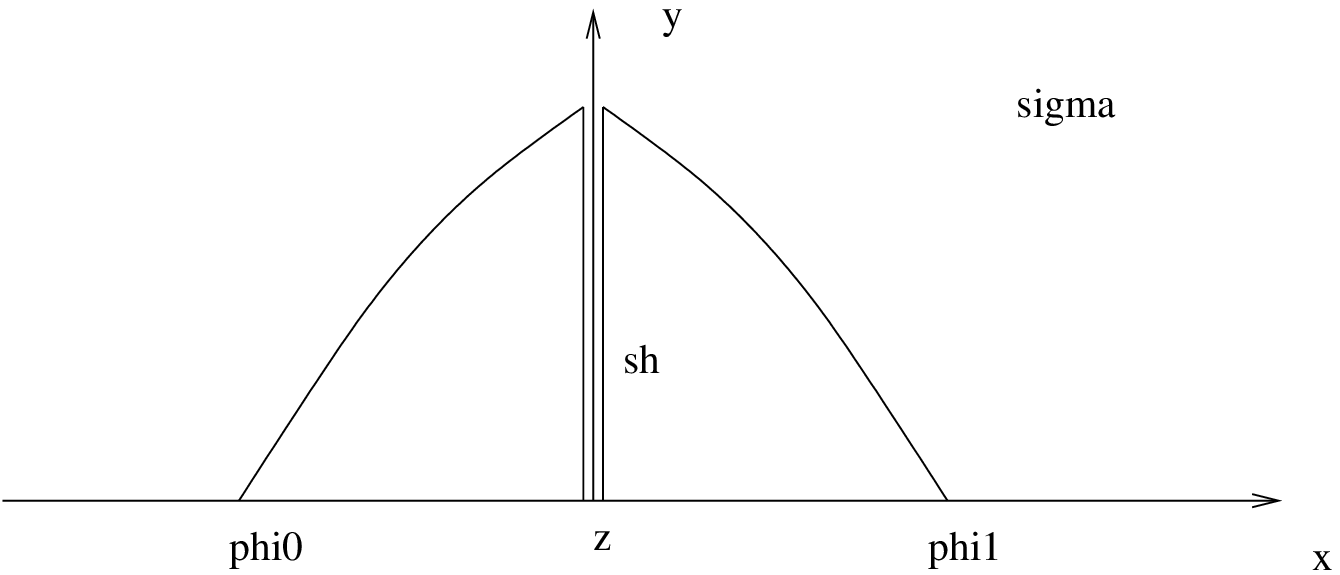}
\end{center}
\caption{}
\end{figure}
For $\sigma = \sigma_- = -1 (\sigma_+ = +1)$ the corresponding
pictures are shown in Figs.4,\, 5 and 6.
\begin{figure}[H]
\unitlength=1in
\begin{center}
\psfragscanon
\psfrag{y}{$\frac{R}{R_0}$}
\psfrag{x}{$\frac{n}{R_0}$}
\psfrag{z}{$0$}
\psfrag{phi0}{$\phi_0-\pi$}
\psfrag{phi1}{$\phi_0$}
\psfrag{sigma}{$\sigma=-1$}
\psfrag{n}{$n<0$}
\includegraphics[width=3.5in]{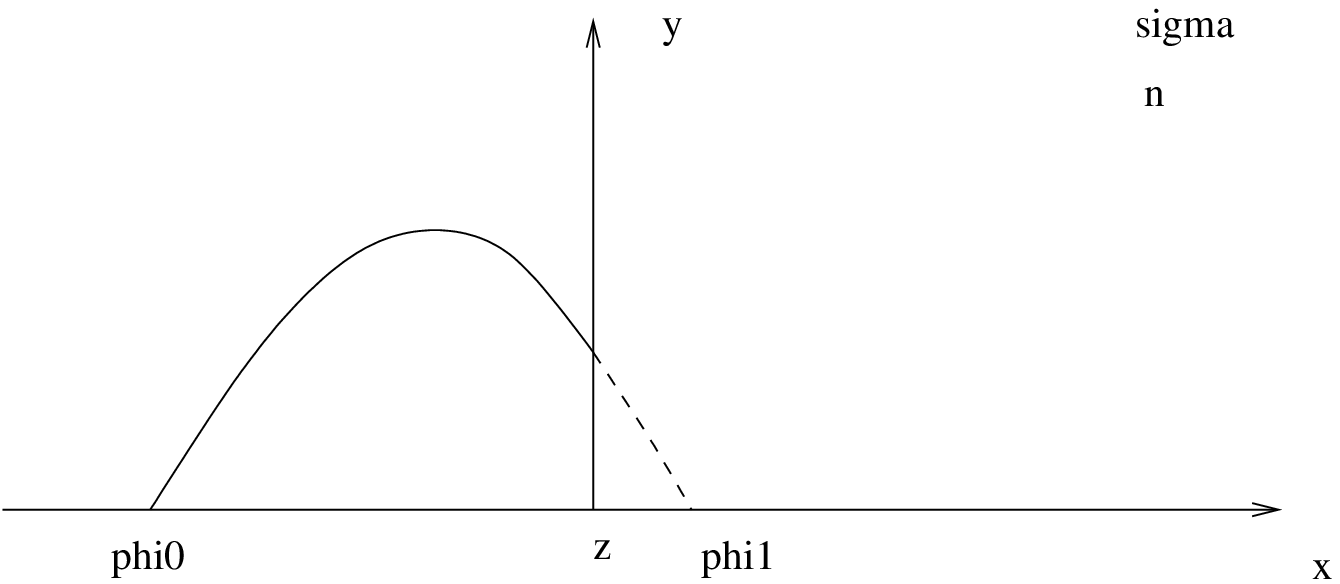}
\end{center}
\caption{}
\end{figure}
\begin{figure}[H]
\unitlength=1in
\begin{center}
\psfragscanon
\psfrag{y}{$\frac{R}{R_0}$}
\psfrag{x}{$\frac{n}{R_0}$}
\psfrag{z}{$0$}
\psfrag{phi0}{$-\phi_0$}
\psfrag{phi1}{$\pi-\phi_0$}
\psfrag{sigma}{$\sigma=-1$}
\psfrag{n}{$n>0$}
\includegraphics[width=3.5in]{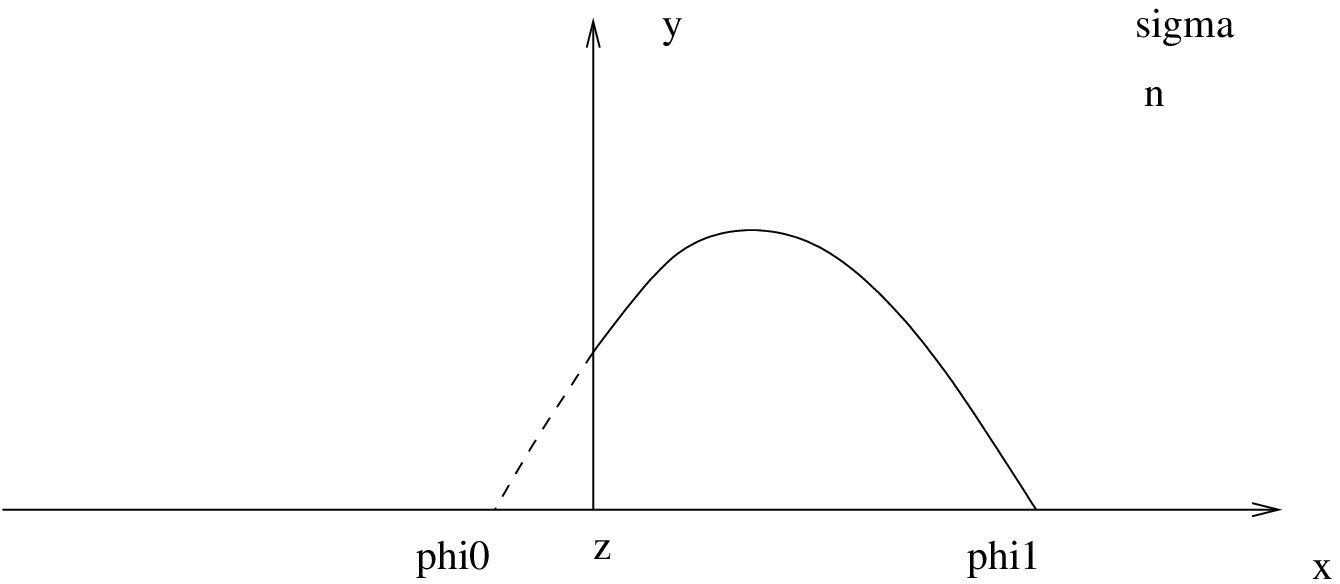}
\end{center}
\caption{}
\end{figure}
\begin{figure}[H]
\unitlength=1in
\begin{center}
\psfragscanon
\psfrag{y}{$\frac{R}{R_0}$}
\psfrag{x}{$\frac{n}{R_0}$}
\psfrag{z}{$0$}
\psfrag{phi0}{$\phi_0-\pi$}
\psfrag{phi1}{$\pi-\phi_0$}
\psfrag{sigma}{$\sigma=-1$}
\psfrag{sh}{$\mbox{shell}$}
\includegraphics[width=3.5in]{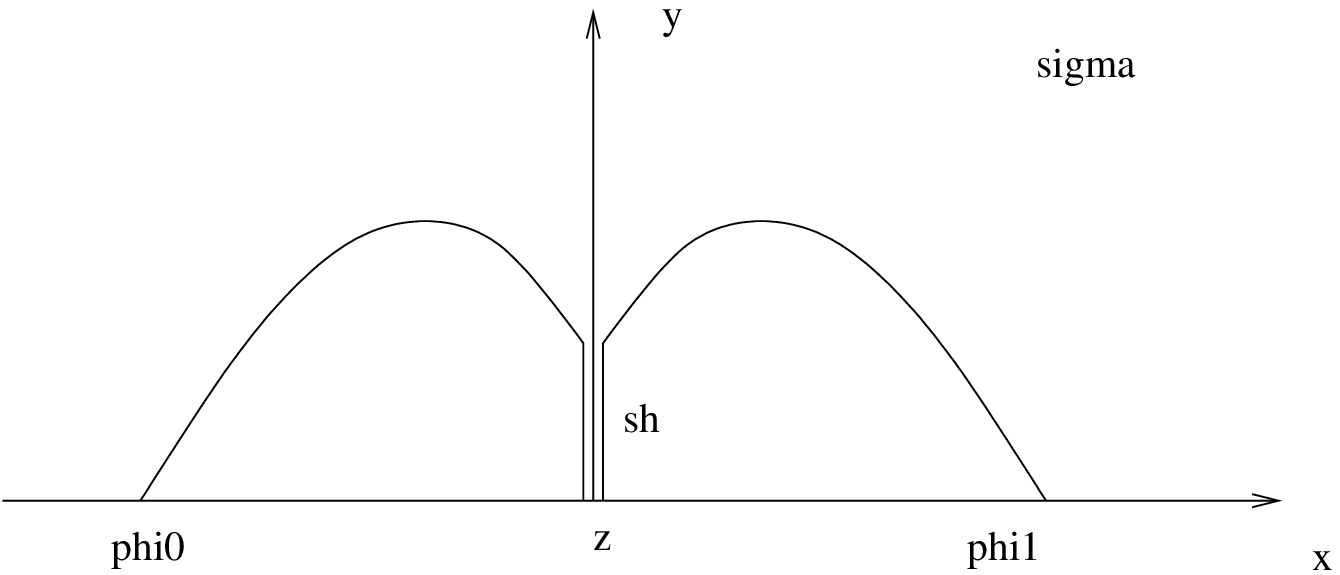}
\end{center}
\caption{}
\end{figure}

Let us now turn to the bulk geometry. The causal structure of
space-times is better seen on the so-called Carter-Penrose conformal
diagrams where each point represents the $(N-1)$ dimensional
homogeneous space. We suppose, everybody in the audience knows how
to construct such a diagram. Below we present only the results
pointing out the $R_{\pm}$- and $T_{\pm}$-regions and corresponding
values of radii $R$ and conformal radii $R^{\star}$ at the
boundaries and horizons. Consider, first, the case $k = +1$. The
relations between $R$ and $R^{\star}$ are now the following
\begin{equation}
\label{RR}
dR^{\star} = \pm \frac{dR}{1 - \frac{R^2}{R^2_0}} \,, \; \Longrightarrow \;\; R^{\star} =
\pm \frac{1}{2} \ln {\frac{1 + \frac{R}{R_0}}{1 - \frac{R}{R_0}}}
\end{equation}
in $R_{\pm}$-regions $0 \le R \le R_0$, and
\begin{equation}
\label{RT}
dR^{\star} = \pm \frac{dR}{\frac{R^2}{R^2_0} - 1} \,, \; \Longrightarrow \;\; R^{\star} =
\pm \frac{1}{2} \ln{\frac{\frac{R}{R_0} + 1}{\frac{R}{R_0} - 1}}
\end{equation}
in $T_{\pm}$-regions, $R_0 \le R \le \infty$. The Carter-Penrose
diagram is the well known square for the de Sitter space-time. The
time coordinate points up, while the radial coordinate goes from
left to right, and the null curves are straight lines with $\pm
45^{\circ}$ (Figs.7 and 8.
\begin{figure}[H]
\unitlength=1in
\begin{center}
\psfragscanon
\psfrag{rz1}{$R=0$}
\psfrag{rz2}{$R=0$}
\psfrag{rinf1}{$R=\infty$}
\psfrag{rinf2}{$R=\infty$}
\psfrag{r01}{$R=R_0$}
\psfrag{r02}{$R=R_0$}
\psfrag{r03}{$R=R_0$}
\psfrag{r04}{$R=R_0$}
\psfrag{rs01}{$R^*=0$}
\psfrag{rs02}{$R^*=0$}
\psfrag{rs03}{$R^*=0$}
\psfrag{rs04}{$R^*=0$}
\psfrag{rsinf1}{$R^*=\infty$}
\psfrag{rsinf2}{$R^*=\infty$}
\psfrag{rsminf1}{$R^*=-\infty$}
\psfrag{rsminf2}{$R^*=-\infty$}
\includegraphics[width=6.0in]{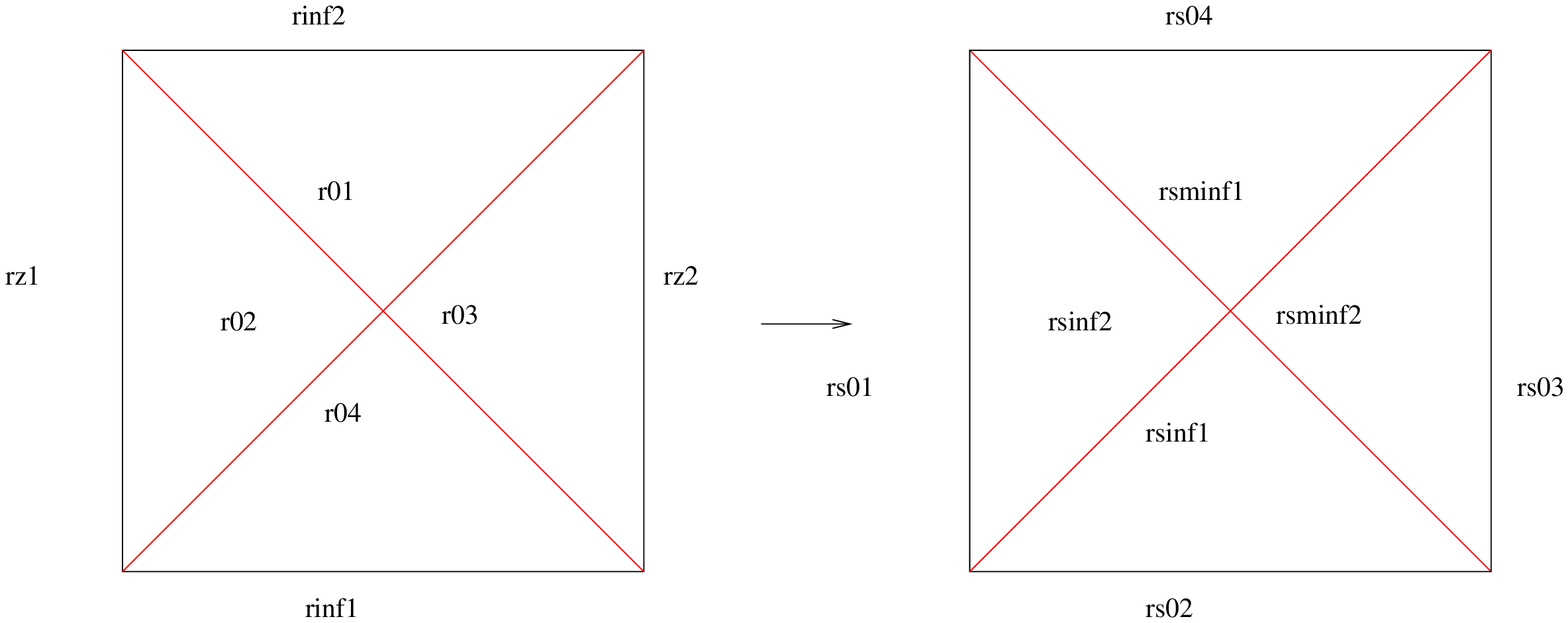}
\end{center}
\caption{}
\end{figure}
\begin{figure}[H]
\unitlength=1in
\begin{center}
\psfragscanon
\psfrag{rp}{$R_+$}
\psfrag{rm}{$R_-$}
\psfrag{tp}{$T_+$}
\psfrag{tm}{$T_-$}
\includegraphics[width=3.5in]{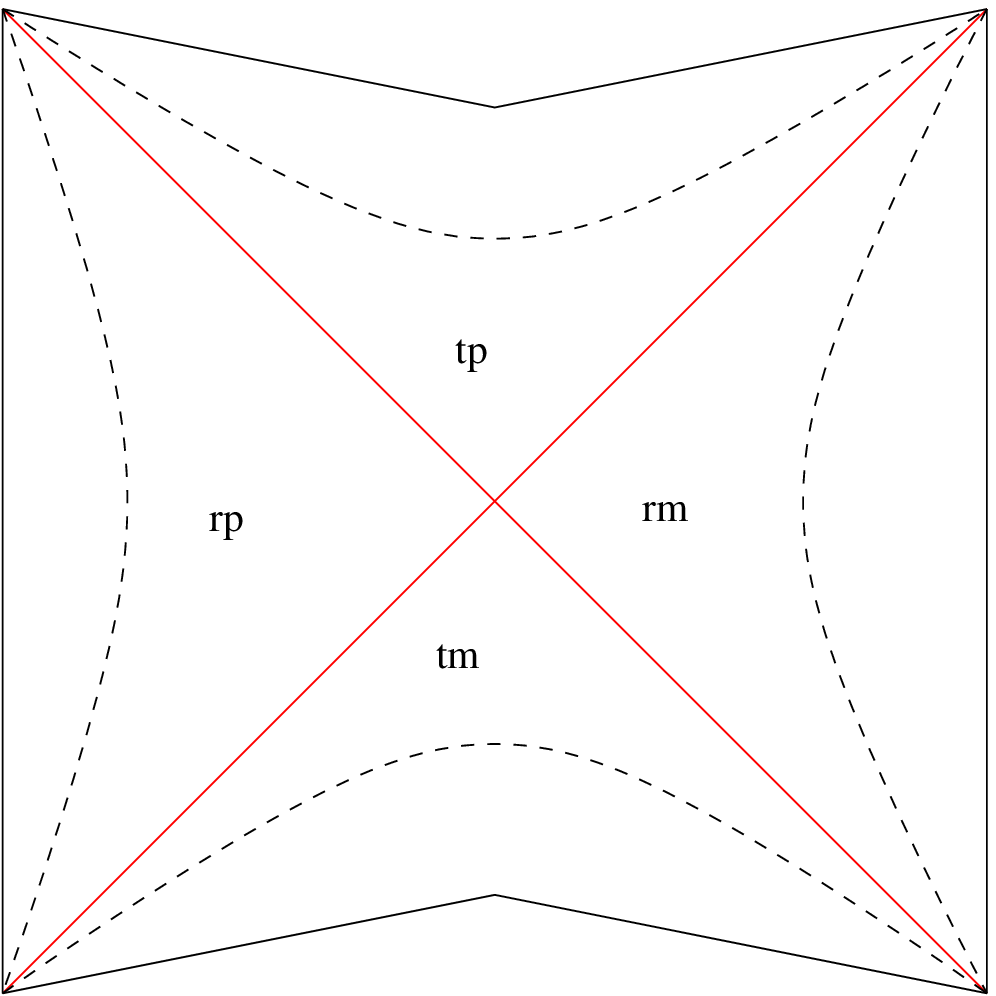}
\end{center}
\caption{}
\end{figure}
In Fig.8 the dashed curves represent the surfaces $R = const$
(time-like in $R$-regions and space-like in $T$-regions), and we
slightly distorted the $R = \infty$ space-like boundaries in order
to make the matchings of the brane to the bulk more visual. And,
finally, the conformal diagrams for the complete geometry of the
space-time with the brane in the case $\Lambda > 0, \, k = +1$ are
shown in Figs.9 and 10. Clearly, they are different for $S^0_0 > 0
\; (\sigma = \sigma_- = +1)$ and for $S^0_0 < 0 \; (\sigma =
\sigma_- = -1)$.
\begin{figure}[H]
\unitlength=1in
\begin{center}
\psfragscanon
\psfrag{rp}{$R_+$}
\psfrag{rm}{$R_-$}
\psfrag{tp1}{$T_+$}
\psfrag{tm1}{$T_-$}
\psfrag{tp2}{$T_+$}
\psfrag{tm2}{$T_-$}
\psfrag{rz1}{$R=0$}
\psfrag{rz2}{$R=0$}
\psfrag{rinf1}{$R=\infty$}
\psfrag{rinf2}{$R=\infty$}
\psfrag{r01}{$R=R_0$}
\psfrag{r02}{$R=R_0$}
\psfrag{r03}{$R=R_0$}
\psfrag{r04}{$R=R_0$}
\includegraphics[width=3.5in]{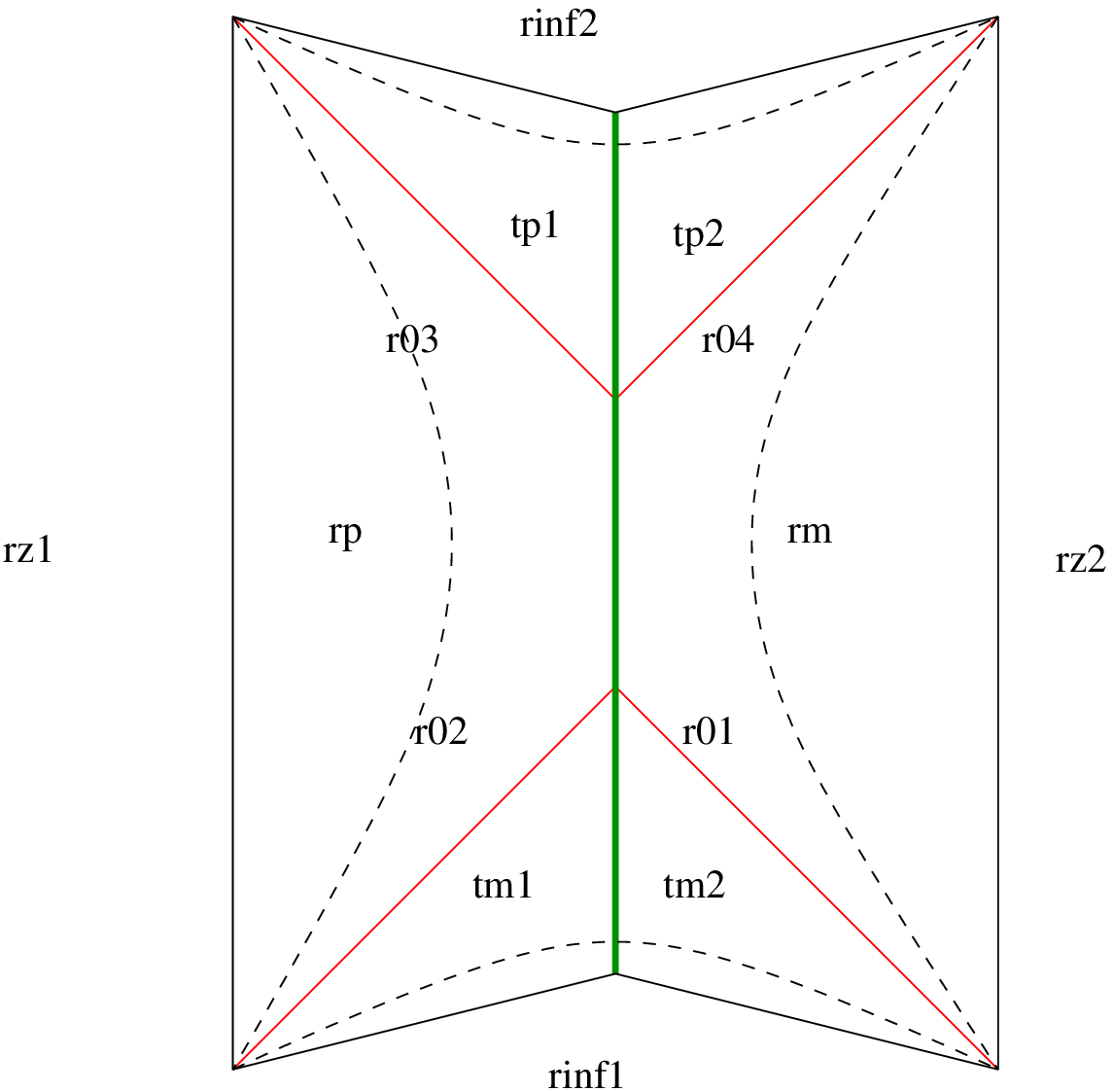}
\end{center}
\caption{}
\end{figure}
\begin{figure}[H]
\unitlength=1in
\begin{center}
\psfragscanon
\psfrag{rp1}{$R_+$}
\psfrag{rm1}{$R_-$}
\psfrag{rp2}{$R_+$}
\psfrag{rm2}{$R_-$}
\psfrag{tp1}{$T_+$}
\psfrag{tm1}{$T_-$}
\psfrag{tp2}{$T_+$}
\psfrag{tm2}{$T_-$}
\psfrag{rinf1}{$R=\infty$}
\psfrag{rinf2}{$R=\infty$}
\psfrag{rinf3}{$R=\infty$}
\psfrag{rinf4}{$R=\infty$}
\psfrag{rz1}{$R=0$}
\psfrag{rz2}{$R=0$}
\includegraphics[width=4.0in]{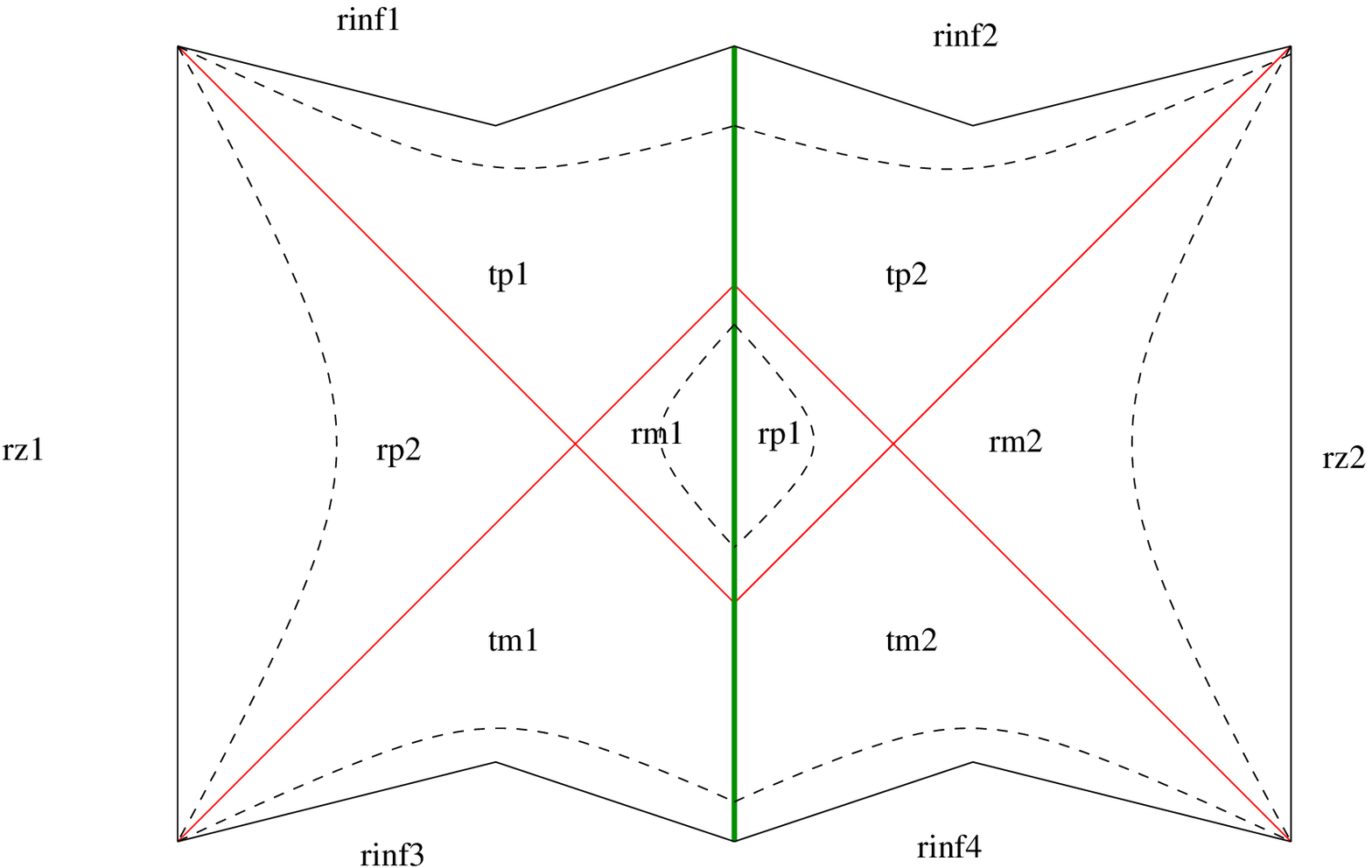}
\end{center}
\caption{}
\end{figure}
Dashed curves are hyper-surfaces $R = const$. The case of negative
surface energy density $S^0_0 < 0$ is much more interesting from the
physics point of view. The bulk geometry on both sides of the brane
has the Einstein-Rosen bridge, or a throat, at the intersection of
horizons $R = R_0$ (this is the so-called bifurcation point). In
this sense such a geometry reminds that of non-traversable wormhole.
The interesting physics begins if there are several more branes
(say, two) in the space-time. Let us imaging that one of the
additional branes is located on the same side of the Einstein-Rosen
bridge as "our" brane is (to the left on the diagram), while the
second one is on the other side (to the right). In classical theory
their existence does not affect the dynamics of "our" shell or
destroy the $Z_2$-symmetry of the matching. But in quantum theory
these additional shells will cause the energy level splitting and
such a splitting will inevitably be asymmetric, resulting in an
asymmetric hierarchy of fundamental interactions.

Consider now the more simple case $k = 0$. Evidently, we have only
the $T_{\pm}$-regions everywhere except the hyper-surfaces $R = 0$
that serve as the apparent horizons. The $R$ and $R^{\star}$
(time-like) are reciprocal,
\begin{equation}
\label{recipr}
R^{\star} = \mp \frac{1}{R}
\end{equation}
in $T_{\pm}$-regions, and the conformal diagrams for the bulk are
simple orthogonal triangles, Fig.11.
\begin{figure}[H]
\unitlength=1in
\begin{center}
\psfragscanon
\psfrag{tp}{$T_+$}
\psfrag{tm}{$T_-$}
\psfrag{rz1}{$R=0$}
\psfrag{rz2}{$R=0$}
\psfrag{rsinf}{$R^*=\infty$}
\psfrag{rsminf}{$R^*=-\infty$}
\psfrag{rinf1}{$R=\infty, R^*=0$}
\psfrag{rinf2}{$R=\infty,R^*=0$}
\includegraphics[width=5.5in]{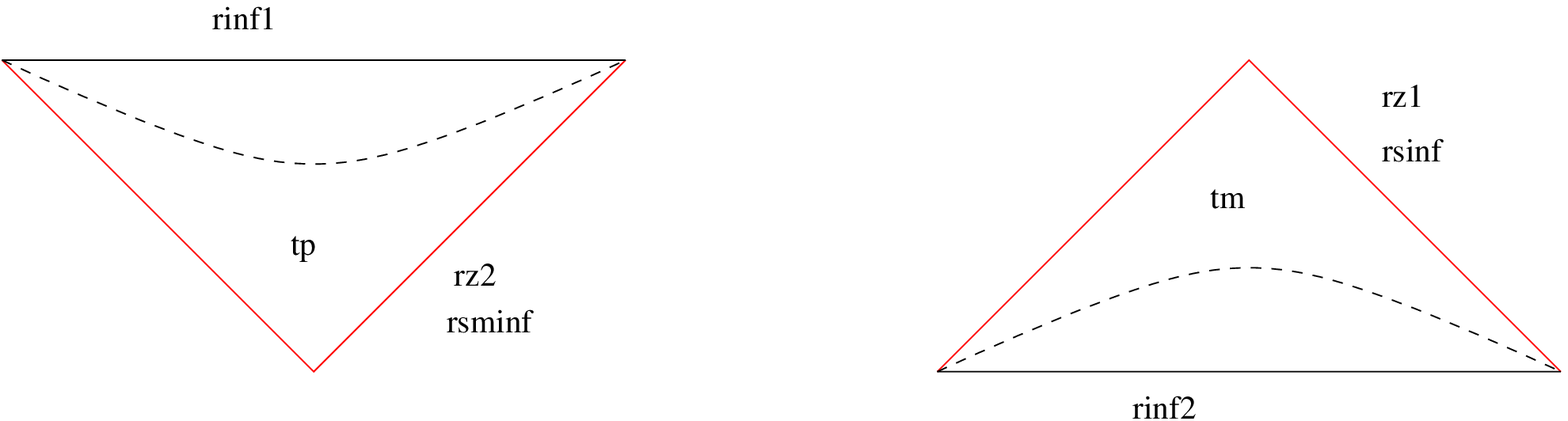}
\end{center}
\caption{}
\end{figure}
The complete geometries with the brane are Figs.12 and 13.
\begin{figure}[H]
\unitlength=1in
\begin{center}
\psfragscanon
\psfrag{rz1}{$R=0$}
\psfrag{rz2}{$R=0$}
\psfrag{rinf}{$R=\infty$}
\includegraphics[width=3.5in]{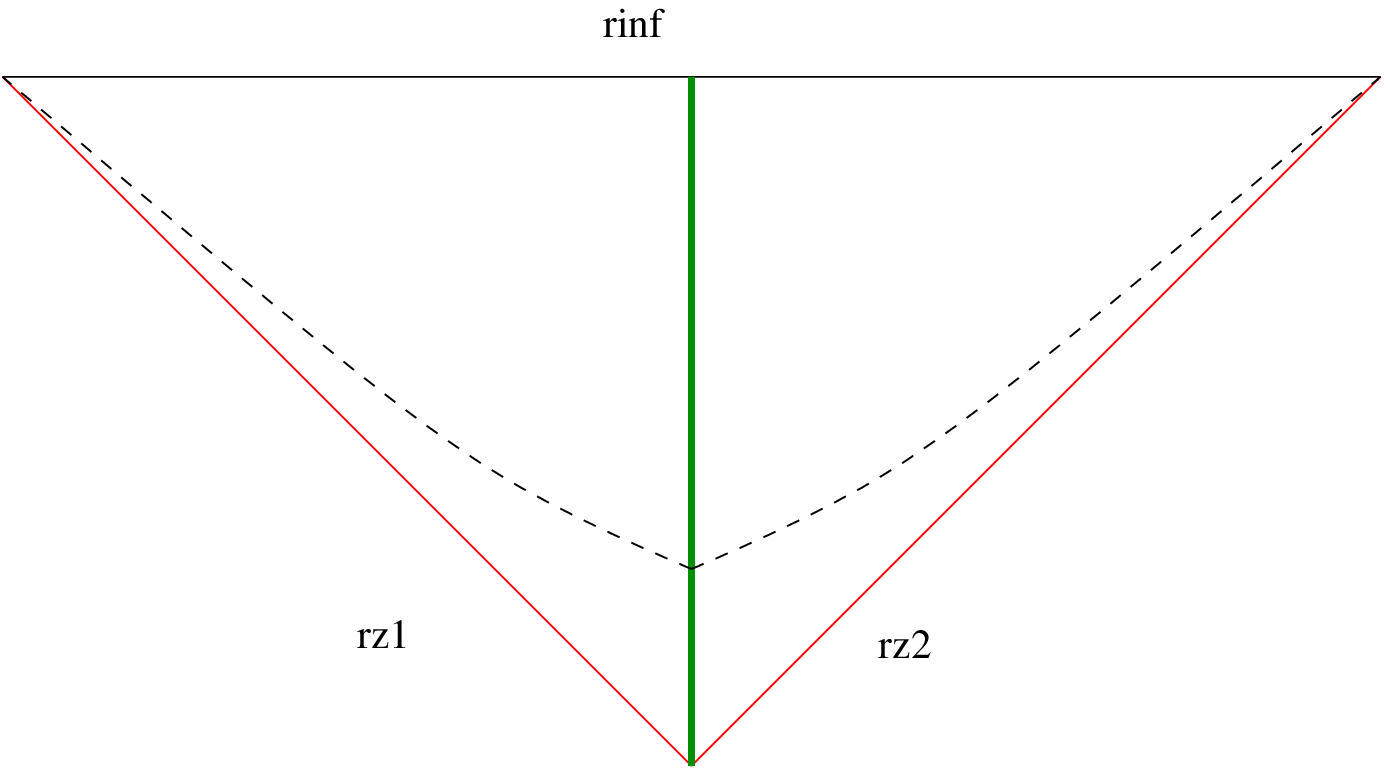}
\end{center}
\caption{}
\end{figure}
\begin{figure}[tbh]
\unitlength=1in
\begin{center}
\psfragscanon
\psfrag{rz1}{$R=0$}
\psfrag{rz2}{$R=0$}
\psfrag{rz3}{$R=0$}
\psfrag{rinf1}{$R=\infty$}
\psfrag{rinf2}{$R=\infty$}
\includegraphics[width=4.0in]{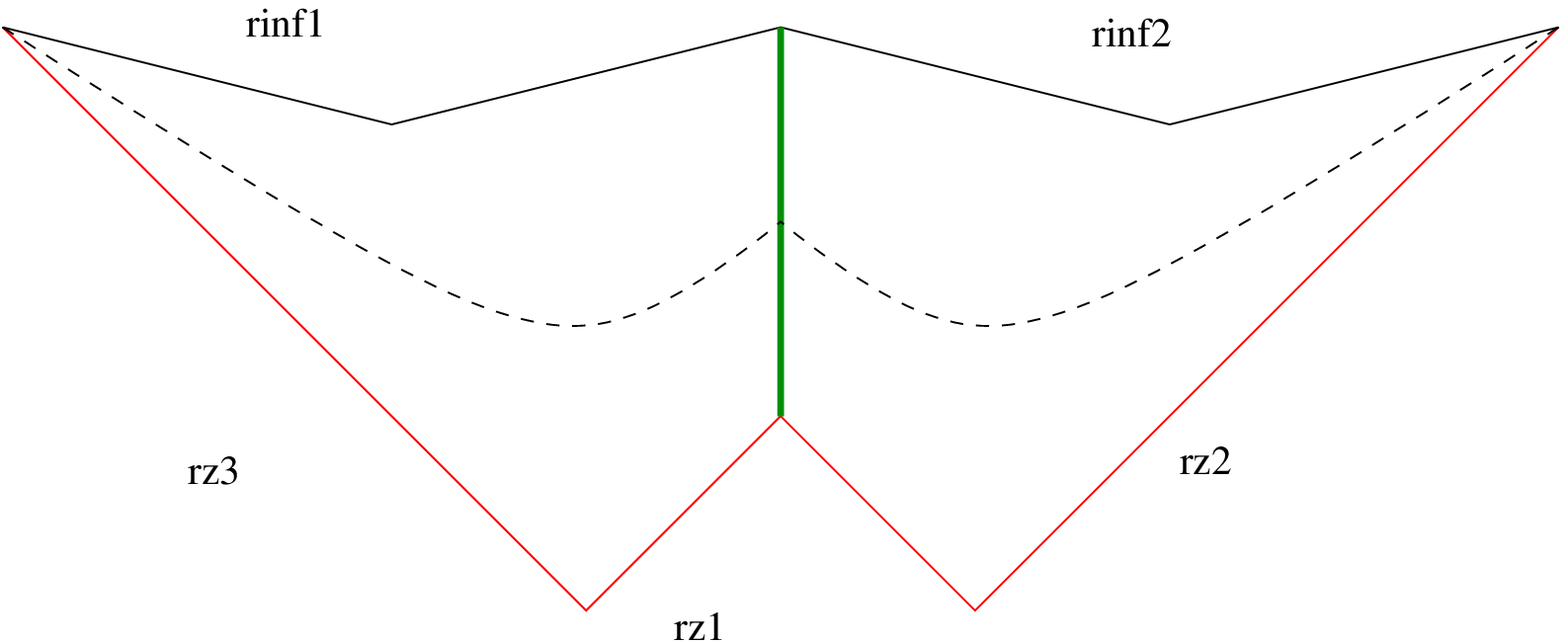}
\end{center}
\caption{}
\end{figure}
plus their time reversals. Again, the dashed curves represent
constant radii.

The case $k = -1$ is a little bit more complex. The whole region $0
\le R < \infty$ is now the $T$-region without horizons. For the
conformal time $R^{\star}$ we have
\begin{equation}
\label{RR-}
R^{\star} = \pm \arctan{\frac{R}{R_0}} \,, \; \Longrightarrow \;\;
R = \pm R_0 \tan{\frac{R^{\star}}{R_0}} \,,
\end{equation}
where the signs $"\pm"$ stand for $T_{\pm}$-regions. when $R$
increases from zero to infinity, $0 \le R < \infty$, the
cosmological time $R^{\star}$ changes from $0$ to $\frac{\pi}{2} \,
R_0$ in $T_+$-region, the region of inevitable expansion (in
$T_-$-region $- \frac{\pi}{2} \, R_0 \le R^{\star} \le 0$ when $R$
decreases from $\infty$ to $0$, this is the case of inevitable
contraction). Formally, we can extend the conformal time $R^{\star}$
to run from $-\infty$ to $\infty$ and, thus, arrive at the so-called
unfolded description. In the purely vacuum space-time there are no
physical observers, but in more realistic brane universe scenarios
everything depends on the physical conditions inside the shell
(possible appearance of real singularities and so on). Further, it
is easy to notice that the two-dimensional metric for $\Lambda < 0,
\, k = +1$ differs from that one for $\Lambda > 0, \, k = -1$ only
by the signature: $(+-) \to (-+)$. This means that the corresponding
Carter-Penrose conformal diagram can be obtained from that of
conventional anti-de Sitter space-time by interchanging $R_{\pm}$-
and $T_{\pm}$-regions,the horizontal lines being replaced by
vertical ones. With this in mind, we get for the bulk geometry
(Fig.14).
\begin{figure}[H]
\unitlength=1in
\begin{center}
\psfragscanon
\psfrag{rz1}{$R=0, R^*=0$}
\psfrag{rz2}{$R=0, R^*=0$}
\psfrag{rinf1}{$R=\infty, R^*=\frac{\pi}{2}R_0$}
\psfrag{rinf2}{$R=\infty, R^*=-\frac{\pi}{2}R_0$}
\psfrag{tp}{$T_+$}
\psfrag{tm}{$T_-$}
\includegraphics[width=6.0in]{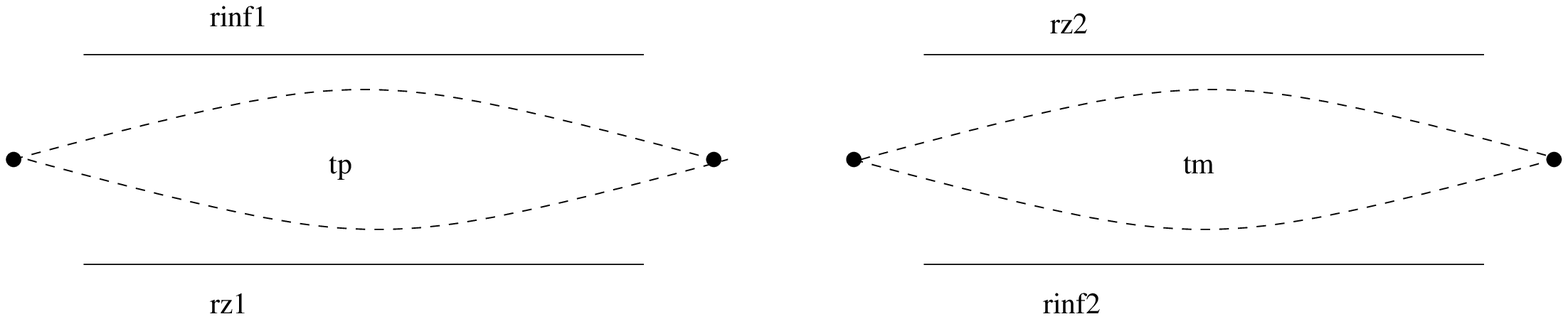}
\end{center}
\caption{}
\end{figure}
Here two isolated points on each diagram are spatial infinities
($-\infty$ on the left and $+\infty$ on the right), and the dashed
curves are for $R = const$. The complete geometries with the brane
for $S^0_0 > 0$ and $S^0_0 < 0$ look on the conformal diagrams as in
Fig.15.
\begin{figure}[H]
\unitlength=1in
\begin{center}
\psfragscanon
\psfrag{rz1}{$R=0, R^*=0$}
\psfrag{rz2}{$R=0, R^*=0$}
\psfrag{rinf1}{$R=\infty, R^*=\frac{\pi}{2}R_0$}
\psfrag{rinf2}{$R=\infty, R^*=-\frac{\pi}{2}R_0$}
\includegraphics[width=6.0in]{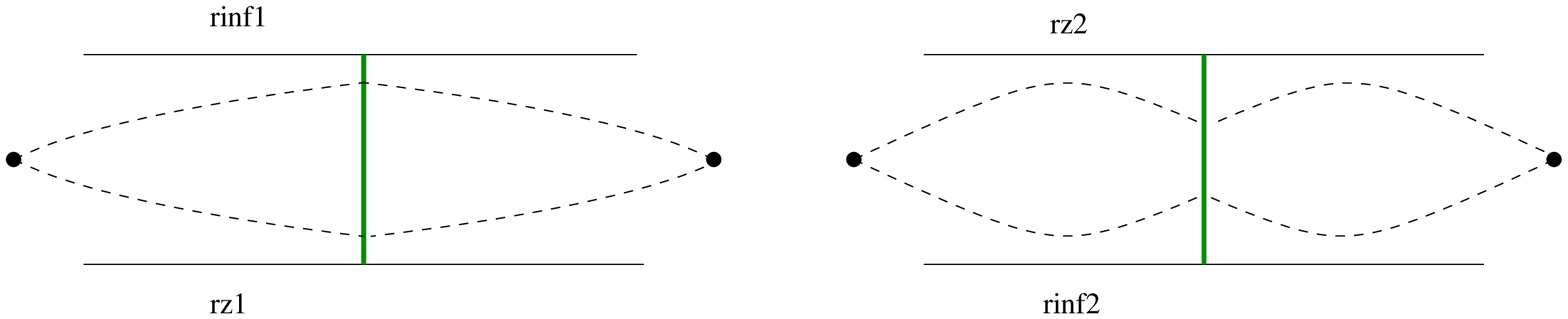}
\end{center}
\caption{}
\end{figure}
This is the case of inevitable expansion. For inevitable contraction
the diagrams are essentially the same.

And now we will describe all possible global geometries when the
bulk is a vacuum $(N+1)$-dimensional space-time with negative
cosmological constant, $\Lambda < 0$. First of all, let us have a
look at the differential equation for $R(n,t)$,
\begin{equation}
\label{Rn1}
R_{,n} = \sigma \sqrt{f^2(t) + k - \frac{2 \Lambda}{N(N-1)} R^2} =
\sigma \sqrt{f^2(t) + k + \frac{R^2}{R^2_0}} \,,
\end{equation}
where we introduced the cosmological radius $R_0 =
\sqrt{\frac{N(N-1)}{2 |\Lambda|}}$. In contrast to the case of
positive $\Lambda$, there are two possibilities: either $f^2(t) + k
> 0$, or $f^2(t) + k < 0$. We start with the case $f^2(t) + k > 0$,
the solution to the Eqn.(\ref{Rn1}) is
\begin{equation}
\label{sRn1}
R = R_0 \sqrt{f^2(t) + k} \, \sinh{\left( \frac{\sigma n}{R_0} + \varphi (t)\right)}\,.
\end{equation}
The Einstein equations on the brane $(n = 0)$ take now the form
\begin{eqnarray}
\label{bk-}
\frac{\sigma n}{R_0} \coth {\varphi(t)} &=&
\frac{4 \pi \, G}{N-1} \, S^0_0 \,, \nonumber \\
\frac{f^2(t) + k}{a^2} = \frac{a^2_{\tau} + k}{a^2} &=&
\frac{2 \Lambda}{N(N-1)} + \left(\frac{4 \pi \, G}{N-1}\right)^2 {S^0_0}^2 =
\frac{1}{R_0^2\,\sinh^2{\varphi (t)}}.
\end{eqnarray}
Again, for vacuum shells $(S^0_0 = S^2_2 = const)$ we have $\varphi
(t) = \varphi _0 = const$. We see that the transition from the
positive cosmological constant to the negative one results in
replacing the trigonometric by corresponding hyperbolic functions.
Moreover, in this case $(f^2(t) + k > 0)$ the induced energy density
inside the shell is positive, so qualitatively, the inner evolution
is exactly the same as for positive $\Lambda$. Since the absolute
value of the surface energy density is bounded from below, $\left|
S^0_0 \right| > \sqrt{\frac{(N-1) \, |\Lambda |}{2 \pi \, G \, N}}$,
we may call such a brane "the heavy shell". The plots of the
functions $R(n)$ for different $\sigma = \pm 1$ are shown in Figs.16
and 17.
\begin{figure}[H]
\unitlength=1in
\begin{center}
\psfragscanon
\psfrag{y}{$\frac{R}{R_0}$}
\psfrag{x}{$\frac{n}{R_0}$}
\psfrag{z}{$0$}
\psfrag{phi0}{$-\phi_0$}
\psfrag{sigma}{$\sigma=+1$}
\psfrag{n}{$n<0$}
\includegraphics[width=3.5in]{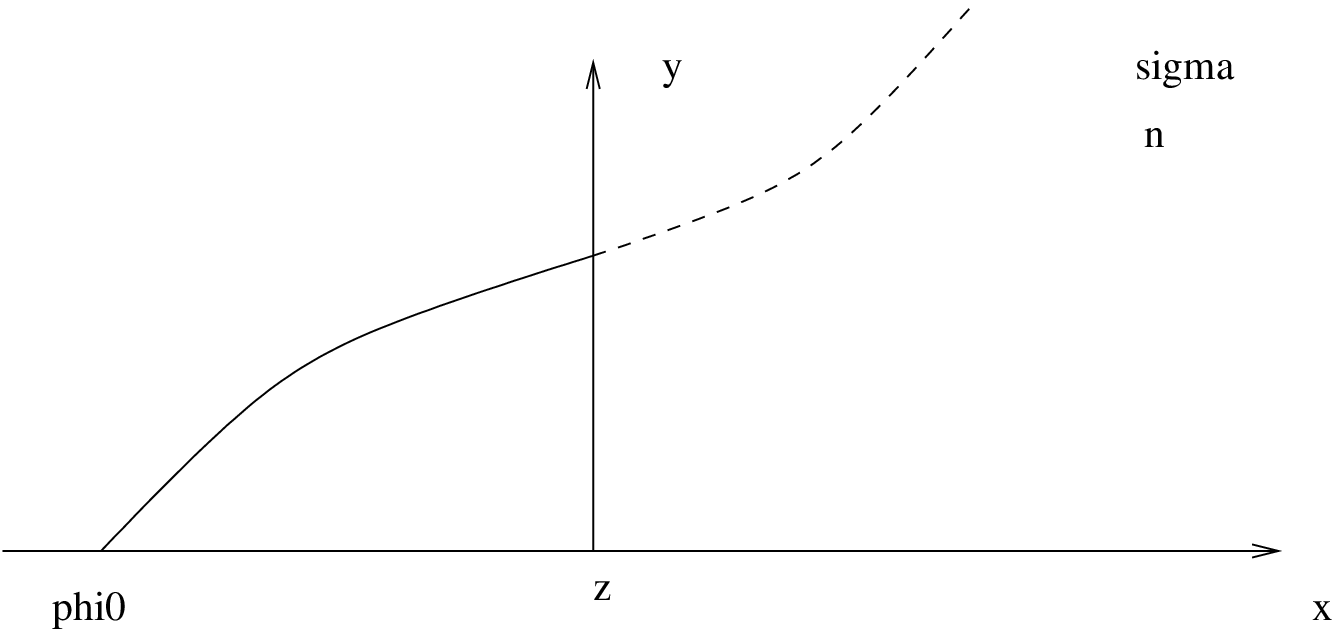}
\end{center}
\caption{}
\end{figure}
\begin{figure}[H]
\unitlength=1in
\begin{center}
\psfragscanon
\psfrag{y}{$\frac{R}{R_0}$}
\psfrag{x}{$\frac{n}{R_0}$}
\psfrag{z}{$0$}
\psfrag{phi0}{$\phi_0$}
\psfrag{sigma}{$\sigma=-1$}
\psfrag{n}{$n<0$}
\includegraphics[width=3.5in]{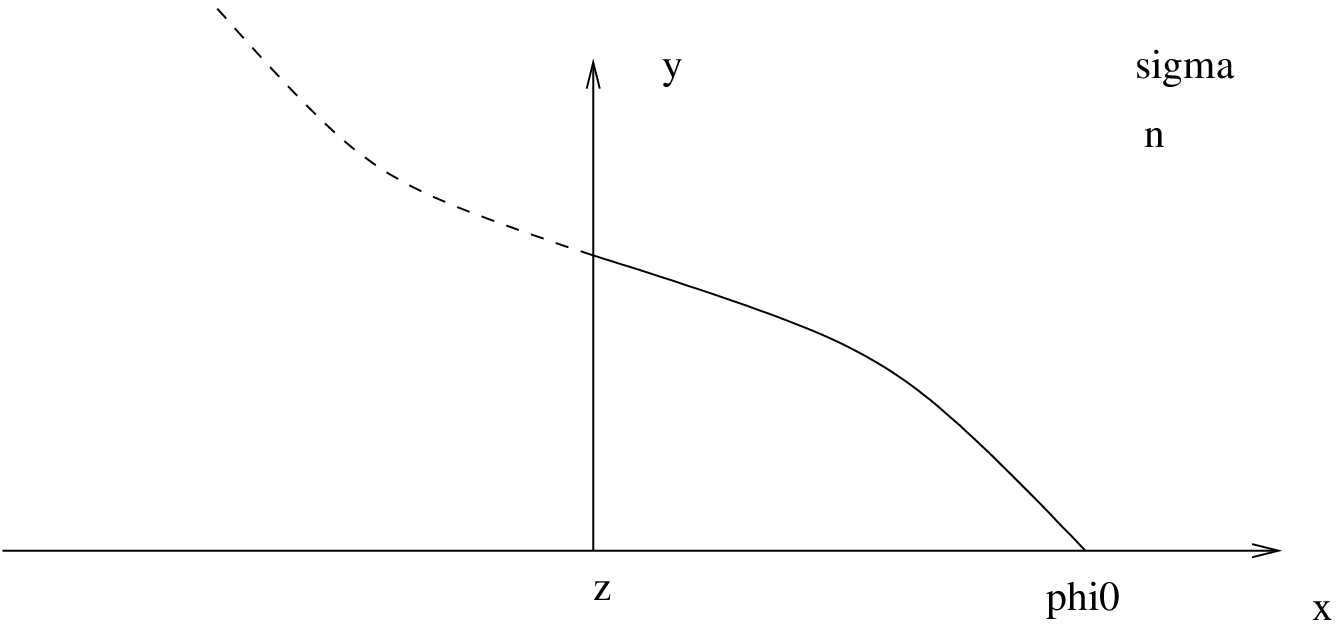}
\end{center}
\caption{}
\end{figure}
With the brane the pictures are Figs.18 and 19.
\begin{figure}[H]
\unitlength=1in
\begin{center}
\psfragscanon
\psfrag{y}{$\frac{R}{R_0}$}
\psfrag{x}{$\frac{n}{R_0}$}
\psfrag{z}{$0$}
\psfrag{phi0}{$-\phi_0$}
\psfrag{phi1}{$\phi_0$}
\psfrag{sigma}{$\sigma=+1$}
\psfrag{sh}{$\mbox{shell}$}
\includegraphics[width=3.5in]{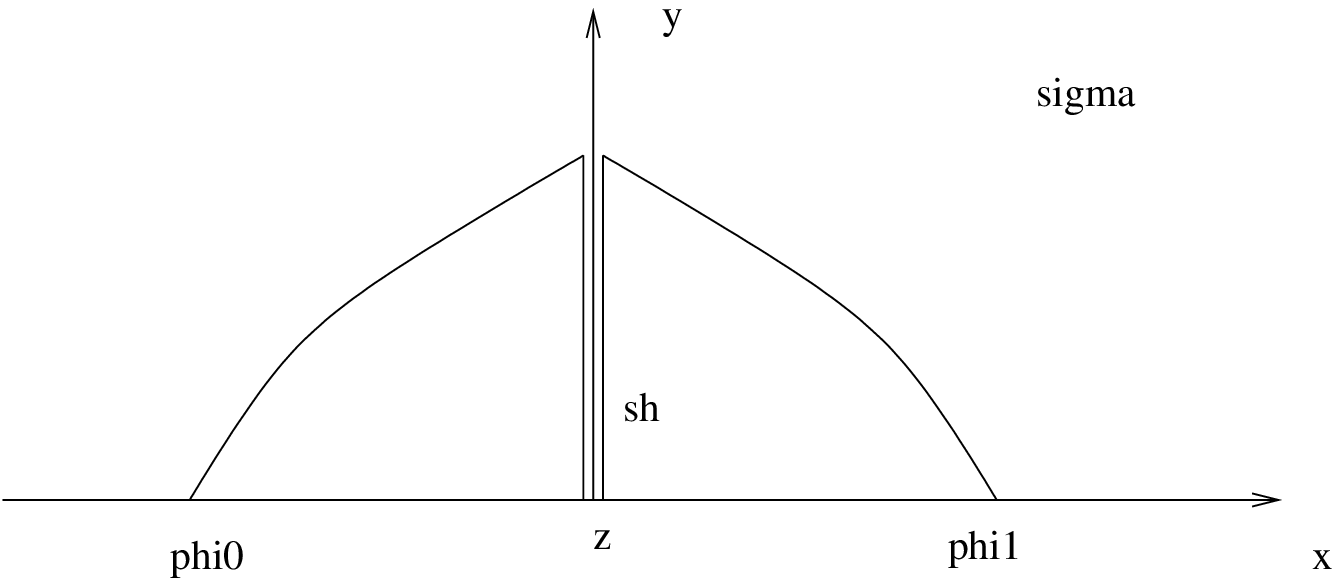}
\end{center}
\caption{}
\end{figure}
\begin{figure}[H]
\unitlength=1in
\begin{center}
\psfragscanon
\psfrag{y}{$\frac{R}{R_0}$}
\psfrag{x}{$\frac{n}{R_0}$}
\psfrag{z}{$0$}
\psfrag{sigma}{$\sigma=-1$}
\psfrag{sh}{$\mbox{shell}$}
\includegraphics[width=3.5in]{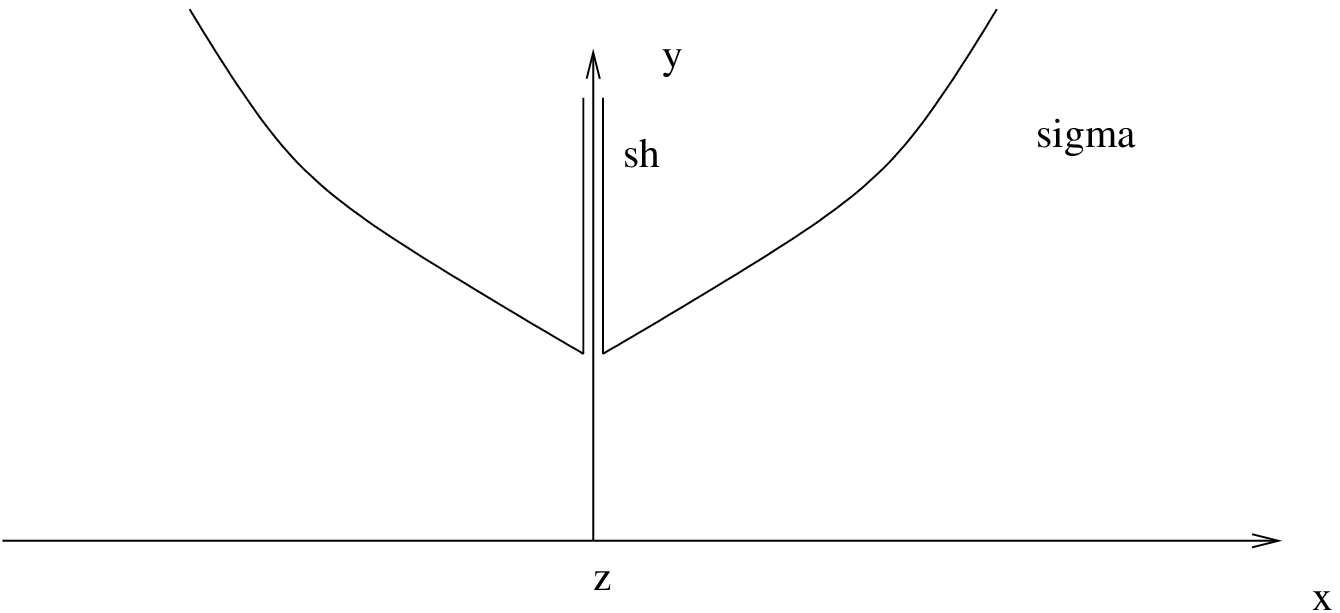}
\end{center}
\caption{}
\end{figure}
Despite of the similar behavior inside the brane, the bulk
geometries are completely different from that for positive
cosmological term.

The case $k = + 1$ is the conventional anti-de Sitter space-time,
and the conformal Carter-Penrose diagram is the same as for $\Lambda
> 0, \, k = - 1$, but it becomes vertical, because instead of
$T$-regions we have now the $R$-regions everywhere, see Fig.20.
\begin{figure}[H]
\unitlength=1in
\begin{center}
\psfragscanon
\psfrag{rinf1}{$R=\infty$}
\psfrag{rinf2}{$R=\infty$}
\psfrag{rs01}{$R^*=\frac{\pi}{2}R_0$}
\psfrag{rs02}{$R^*=-\frac{\pi}{2}R_0$}
\psfrag{rz1}{$R=0$}
\psfrag{rz2}{$R=0$}
\psfrag{rsz1}{$R^*=0$}
\psfrag{rsz2}{$R^*=0$}
\psfrag{rp}{$R_+$}
\psfrag{rm}{$R_-$}
\includegraphics[width=5.5in]{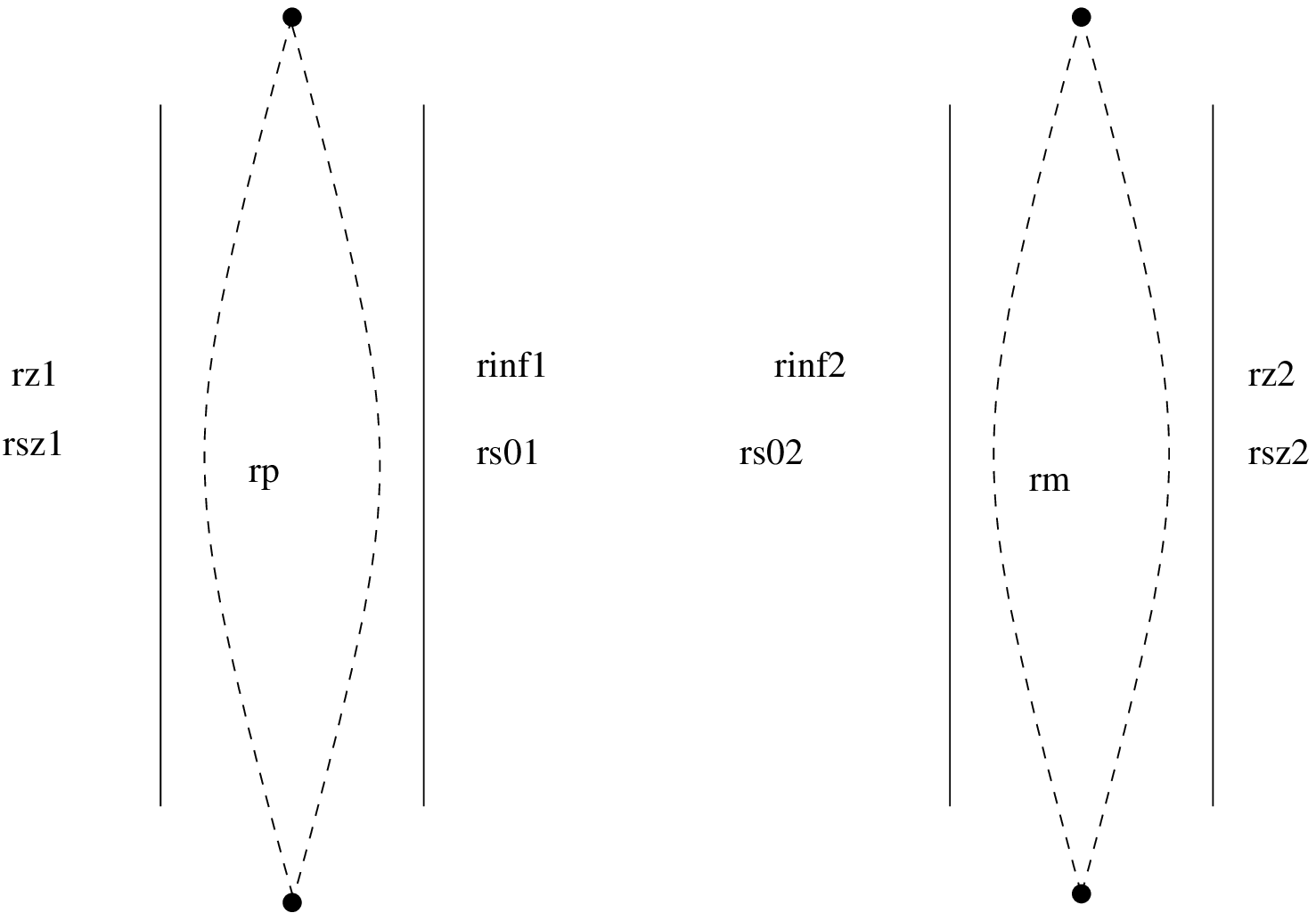}
\end{center}
\caption{}
\end{figure}
The isolated points are the future and past time infinities. Of
course, we can consider a variety of unfolded version (with
different identifications) of this AdS space-time, and such
extensions are even more natural than before, because due to the
negative curvature the light rays reach the "boundary" at $R =
\infty$ in finite coordinate time interval. But, again, everything
depends on the specific properties of the matter inside the brane.
Remembering that the scale factor of the brane evolution for $k = +
1$ is bounded from below we obtain the following two types of global
geometries for $S^0_0 > 0$ and $S_0^0 < 0$, up to possible
unfoldings, Figs.21 and 22.
\begin{figure}[H]
\unitlength=1in
\begin{center}
\psfragscanon
\psfrag{rz1}{$R=0$}
\psfrag{rz2}{$R=0$}
\includegraphics[width=2.5in]{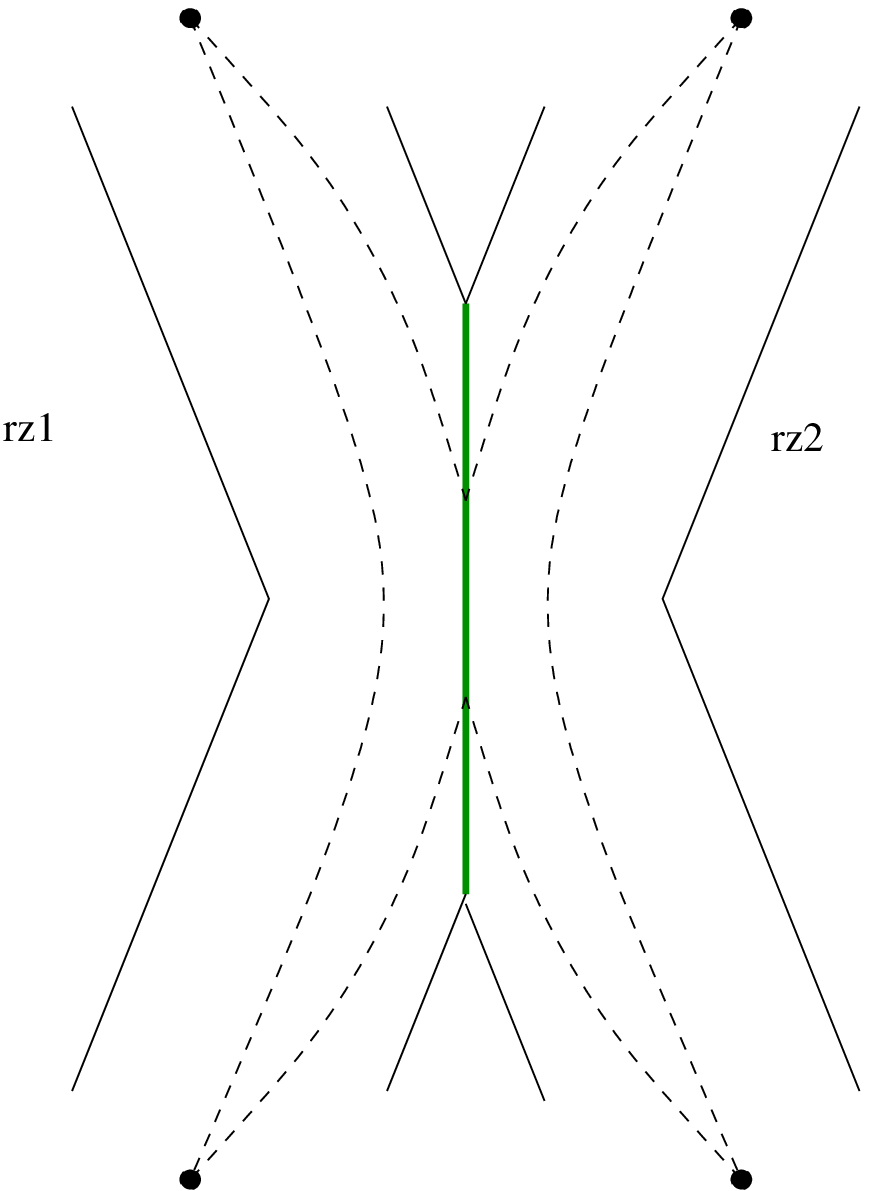}
\end{center}
\caption{}
\end{figure}
\begin{figure}[H]
\unitlength=1in
\begin{center}
\psfragscanon
\psfrag{rp1}{$R_+$}
\psfrag{rp2}{$R_+$}
\psfrag{rm1}{$R_-$}
\psfrag{rm2}{$R_-$}
\includegraphics[width=3.5in]{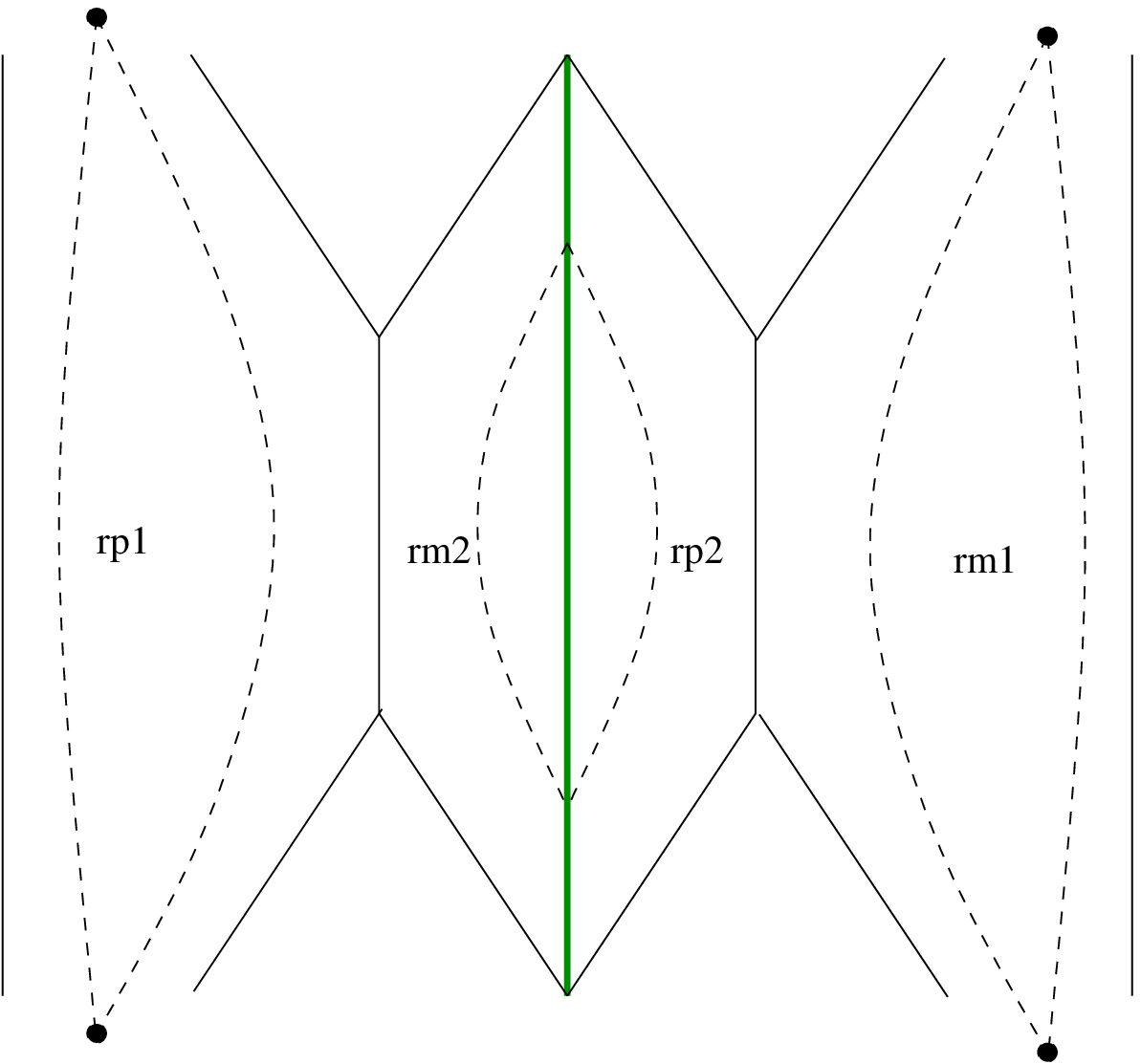}
\end{center}
\caption{}
\end{figure}
The dashed curves represent the hyper-surfaces $R = const$.

In the case $k = 0, \, \Lambda < 0$ everything is similar to that
for $k = 0, \, \Lambda > 0$. Again, the $T$-region is replaced by
the $R$-region, and the vertical orthogonal triangle becomes
horizontal. The surfaces $R = 0$ are the apparent horizons. For the
bulk geometry we have Fig.23.
\begin{figure}[H]
\unitlength=1in
\begin{center}
\psfragscanon
\psfrag{rp}{$R_+$}
\psfrag{rm}{$R_-$}
\psfrag{rz1}{$R=0$}
\psfrag{rz2}{$R=0$}
\psfrag{rsz1}{$R^*=0$}
\psfrag{rsz2}{$R^*=0$}
\psfrag{rsminf}{$R^*=-\infty$}
\psfrag{rsinf}{$R^*=\infty$}
\psfrag{rinf1}{$R=\infty$}
\psfrag{rinf2}{$R=\infty$}
\includegraphics[width=4.5in]{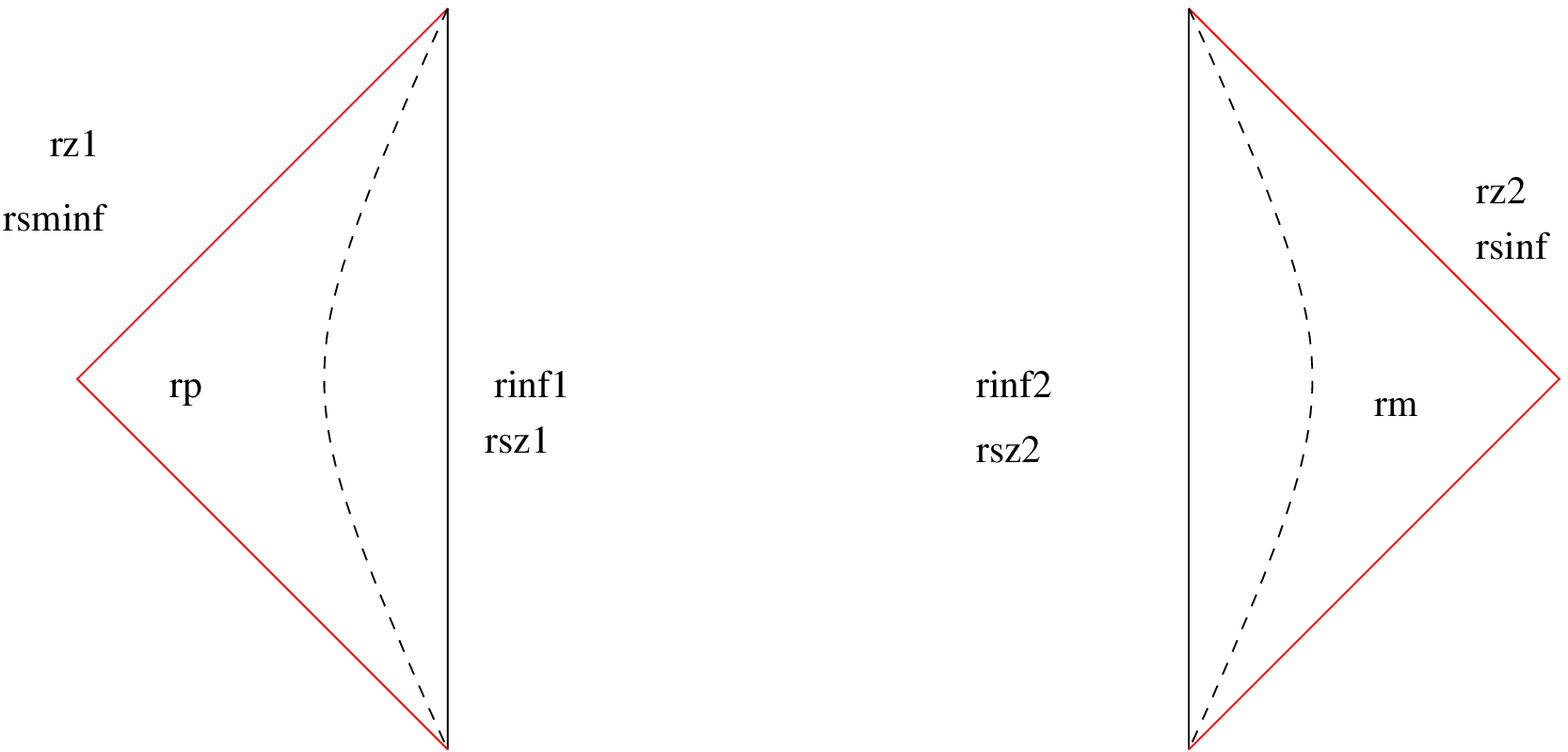}
\end{center}
\caption{}
\end{figure}
After inclusion of the brane we get, Figs.24 and 25.
\begin{figure}[H]
\unitlength=1in
\begin{center}
\psfragscanon
\psfrag{rp}{$R_+$}
\psfrag{rm}{$R_-$}
\psfrag{rz1}{$R=0$}
\psfrag{rz2}{$R=0$}
\psfrag{rz3}{$R=0$}
\psfrag{rz4}{$R=0$}
\psfrag{rinf1}{$R=\infty$}
\includegraphics[width=4.5in]{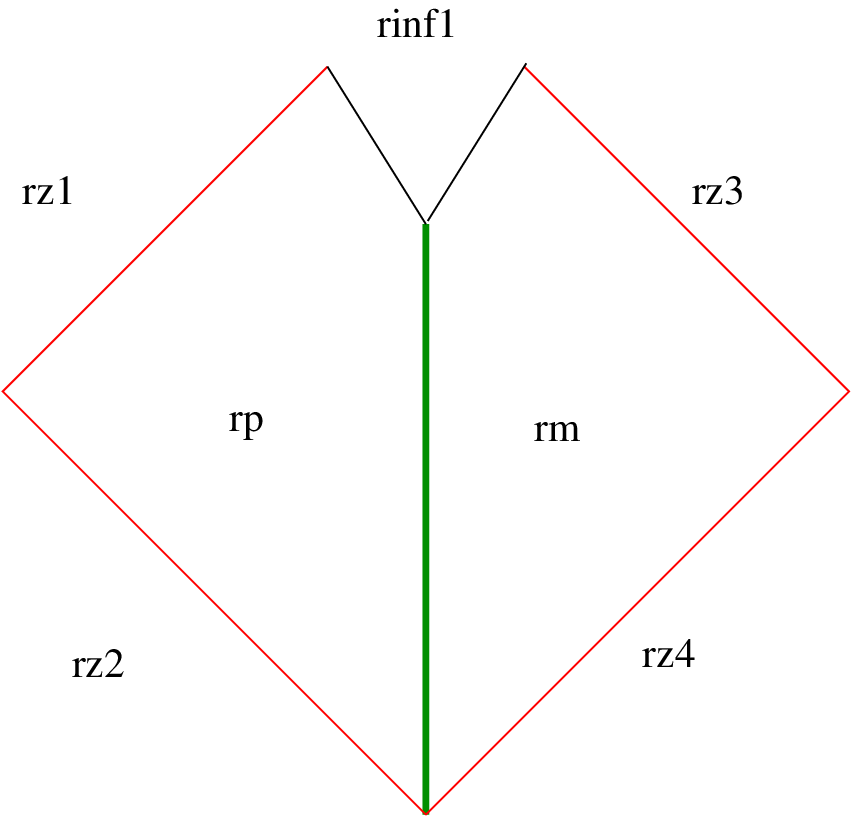}
\end{center}
\caption{}
\end{figure}

\begin{figure}[H]
\unitlength=1in
\begin{center}
\psfragscanon
\psfrag{rp}{$R_+$}
\psfrag{rm}{$R_-$}
\psfrag{rinf1}{$R=\infty$}
\psfrag{rinf2}{$R=\infty$}
\psfrag{rinf3}{$R=\infty$}
\psfrag{rinf4}{$R=\infty$}
\psfrag{rz}{$R=0$}
\includegraphics[width=2.5in]{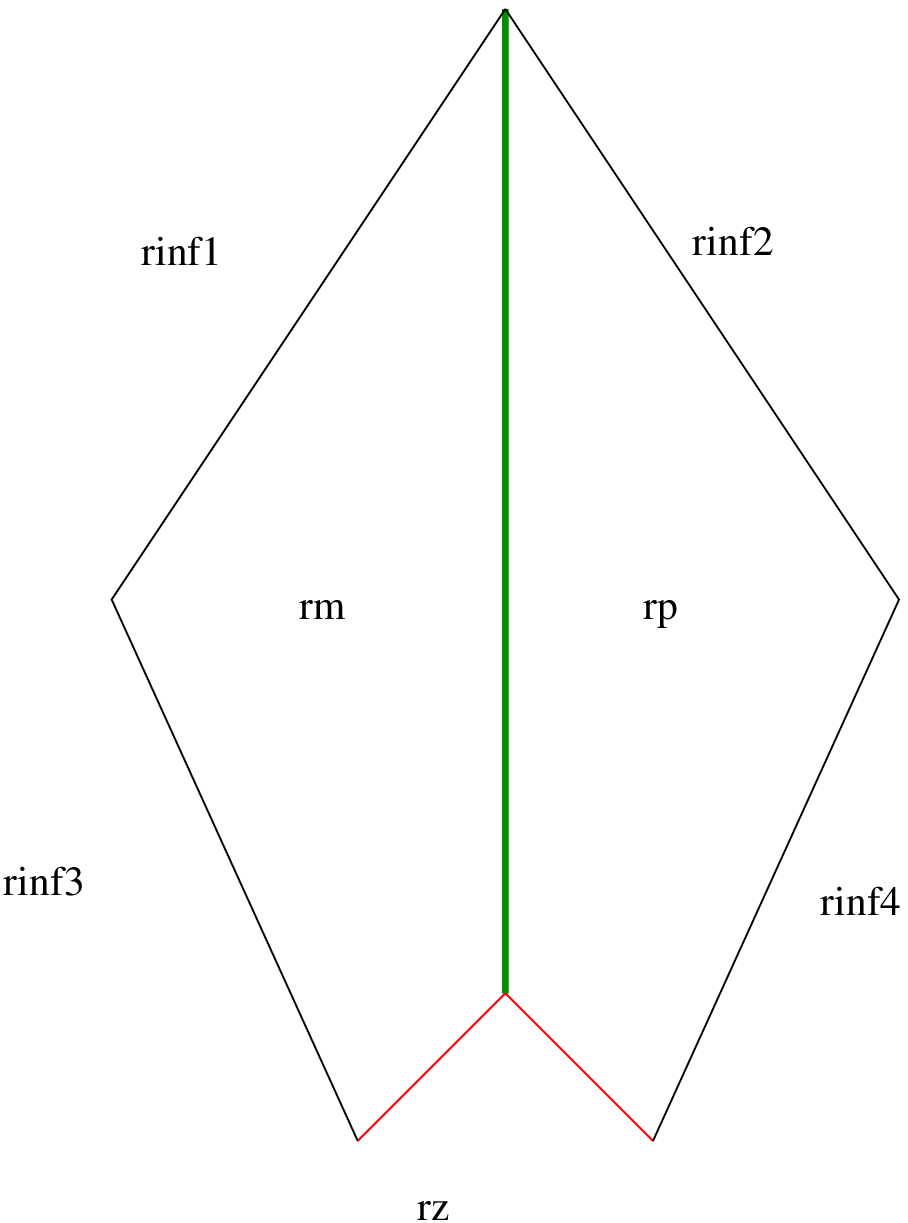}
\end{center}
\caption{}
\end{figure}

The most unusual is the case $\Lambda < 0, \, k = - 1$. It admits
both the "heavy shells" with $\left| S^0_0\right| >
\sqrt{\frac{(N-1) |\Lambda|}{2 \pi \, G N}}$, and the "light shells"
for which $\left| S^0_0\right| < \sqrt{\frac{(N-1) \, |\Lambda |}{2
\pi \, G N}}$. For the "heavy" shells
\begin{equation}
\label{hs}
R = R_0 \, \sinh{\left( \frac{t}{R_0 \sinh{\varphi_0}}\right)} \,
\sinh{\left( \frac{\sigma n}{R_0} + \varphi_0\right)} \,,
\end{equation}
and the shell infinitely expands from zero radius to infinity. For
the "light" shells
\begin{eqnarray}
\label{ls}
R &=& R_0 \, \sin{\left(\frac{t}{R_0 \cosh{\varphi_0}}\right)} \,
\cosh{\left( \frac{\sigma n}{R_0} + \varphi_0\right)}\, ,\nonumber \\
R &=& \pm R_0 \tanh{\frac{R^{\star}}{R_0}} \, ,\: \: \: \: \: \: ,
0 \le R \le R_0 \, , \nonumber\\
R &=& R_0 \coth{\frac{R^{\star}}{R_0}} \, , \: \: \: \: \: \:
R_0 \le R < \infty \, .
\end{eqnarray}
The "light" shells, first expands from zero radius $R = 0$ to the
maximum at $R = R_0 \, \cosh{\varphi_0}$ and then contracts back to
$R = 0$. The curves $R(n,)$ for $t = const$ look in this case as
shown in Figs.26 and 27.

\begin{figure}[H]
\unitlength=1in
\begin{center}
\psfragscanon
\psfrag{y}{$\frac{R}{R_0}$}
\psfrag{x}{$\frac{n}{R_0}$}
\psfrag{z}{$0$}
\psfrag{phi0}{$-\phi_0$}
\psfrag{sigma}{$\sigma=+1$}
\psfrag{n}{$n<0$}
\includegraphics[width=3.5in]{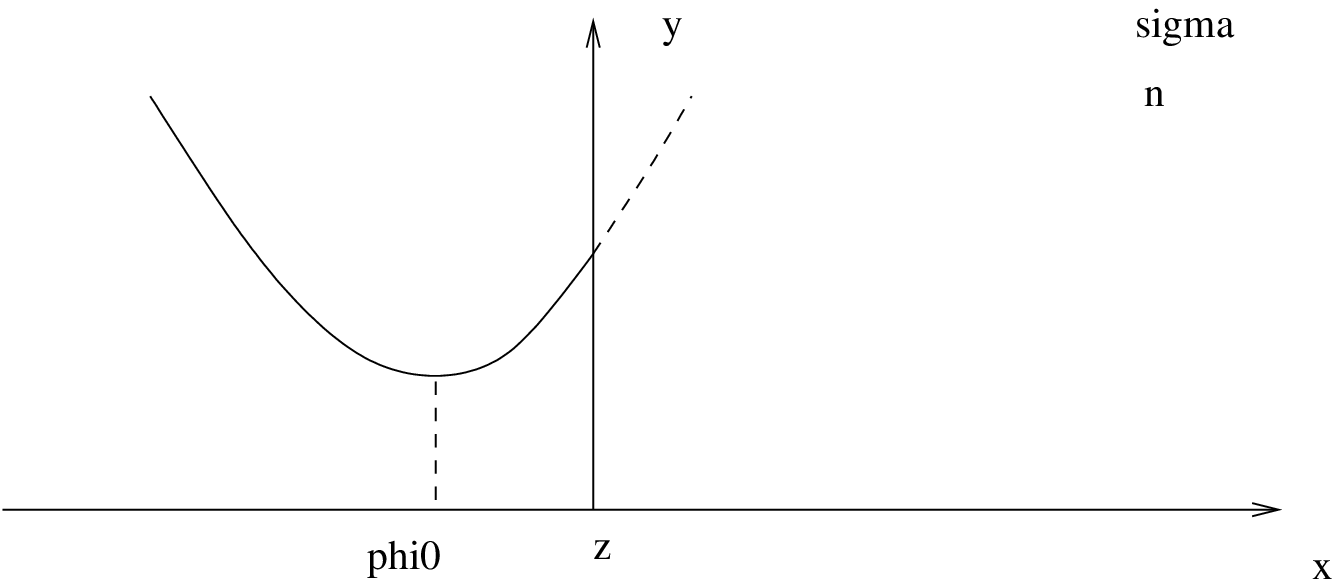}
\end{center}
\caption{}
\end{figure}

\begin{figure}[H]
\unitlength=1in
\begin{center}
\psfragscanon
\psfrag{y}{$\frac{R}{R_0}$}
\psfrag{x}{$\frac{n}{R_0}$}
\psfrag{z}{$0$}
\psfrag{phi0}{$\phi_0$}
\psfrag{sigma}{$\sigma=-1$}
\psfrag{n}{$n>0$}
\includegraphics[width=3.5in]{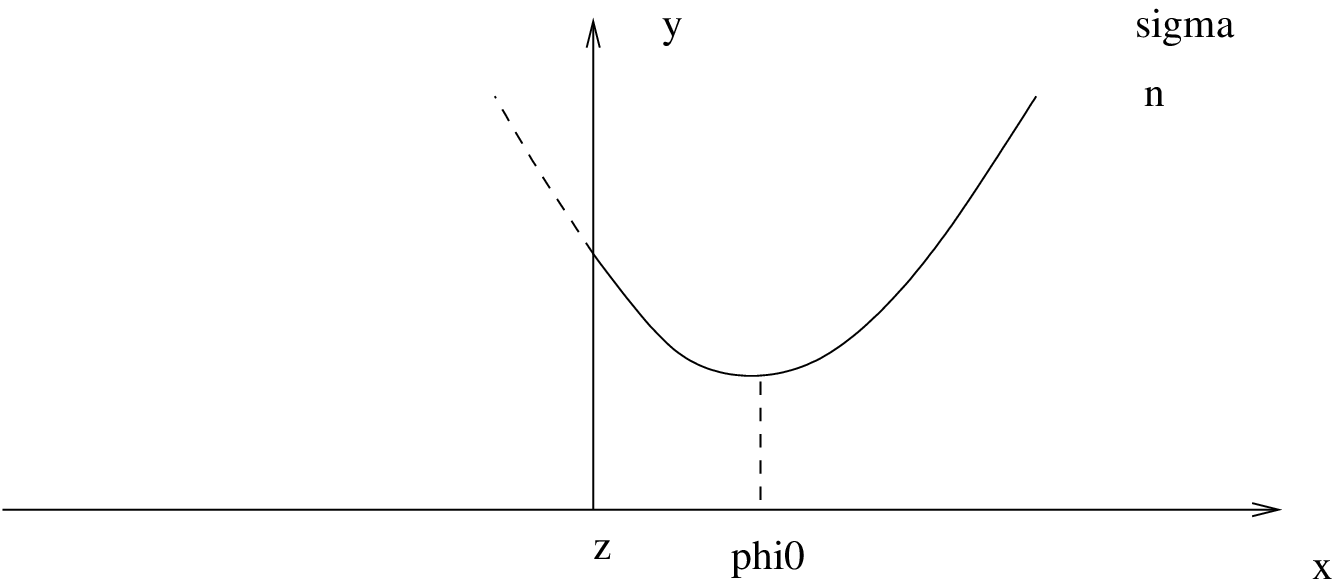}
\end{center}
\caption{}
\end{figure}
With inclusion of the shell, the complete pictures are ($\sigma = +
1$ for $S^0_0 > 0, \, \sigma = - 1$ for $S^0_0 < 0$) Figs.28 and 29.

\begin{figure}[H]
\unitlength=1in
\begin{center}
\psfragscanon
\psfrag{y}{$\frac{R}{R_0}$}
\psfrag{x}{$\frac{n}{R_0}$}
\psfrag{z}{$0$}
\psfrag{phi0}{$-\phi_0$}
\psfrag{phi1}{$\phi_0$}
\psfrag{sigma}{$\sigma=+1$}
\psfrag{sh}{$\mbox{shell}$}
\includegraphics[width=4.5in]{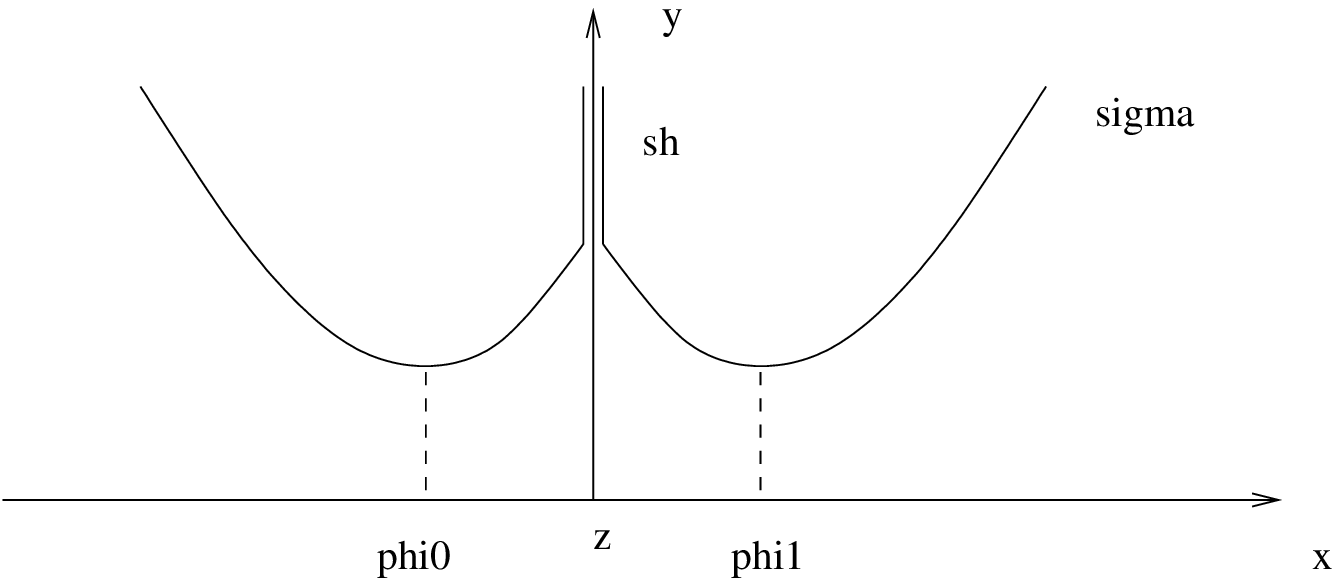}
\end{center}
\caption{}
\end{figure}

\begin{figure}[H]
\unitlength=1in
\begin{center}
\psfragscanon
\psfrag{y}{$\frac{R}{R_0}$}
\psfrag{x}{$\frac{n}{R_0}$}
\psfrag{z}{$0$}
\psfrag{phi0}{$-\phi_0$}
\psfrag{phi1}{$\phi_0$}
\psfrag{sigma}{$\sigma=-1$}
\psfrag{sh}{$\mbox{shell}$}
\includegraphics[width=4.5in]{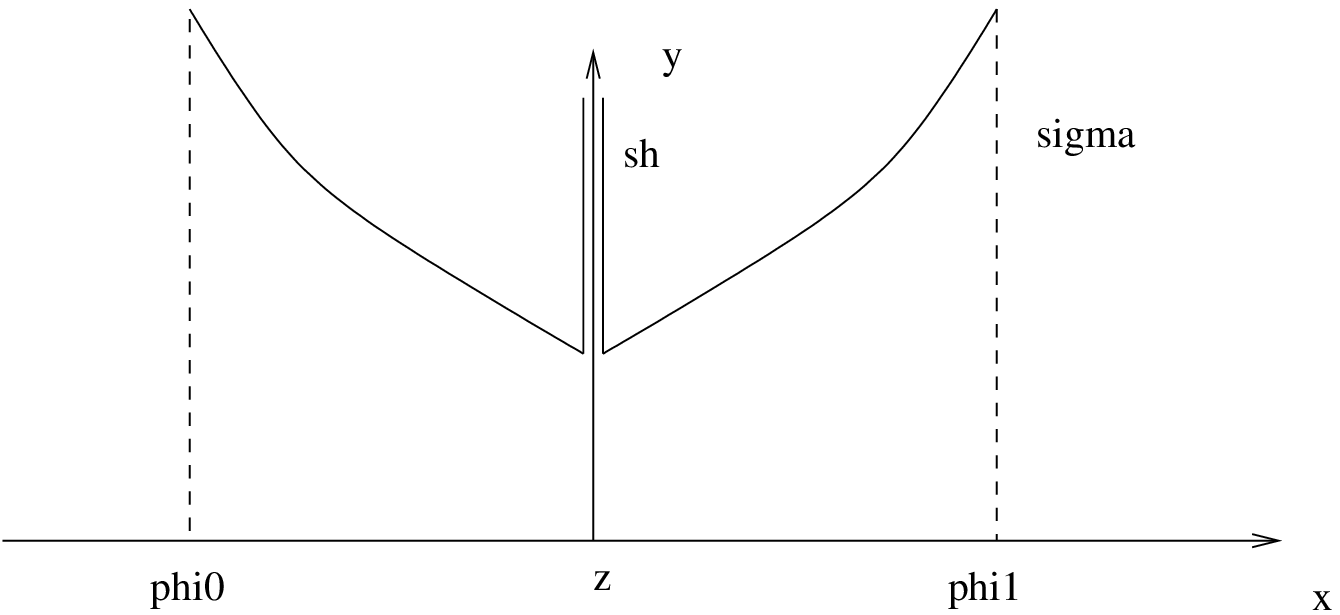}
\end{center}
\caption{}
\end{figure}
The Carter-Penrose diagram is also unusual. Formally, it can be
obtained from that of $\Lambda > 0, \, k = +1$ by interchanging $R$-
and $T$-regions. We get Fig.30.

\begin{figure}[H]
\unitlength=1in
\begin{center}
\psfragscanon
\psfrag{rp}{$R_+$}
\psfrag{rm}{$R_-$}
\psfrag{tp}{$T_+$}
\psfrag{tm}{$T_-$}
\psfrag{rz1}{$R=0, R^*=0$}
\psfrag{rz2}{$R=0, R^*=0$}
\psfrag{rsz1}{$R^*=0$}
\psfrag{rsz2}{$R^*=0$}
\psfrag{rinf1}{$R=\infty$}
\psfrag{rinf2}{$R=\infty$}
\psfrag{rsminf1}{$R^*=-\infty$}
\psfrag{rsminf2}{$R^*=-\infty$}
\psfrag{rsinf1}{$R^*=\infty$}
\psfrag{rsinf2}{$R^*=\infty$}
\includegraphics[width=4.5in]{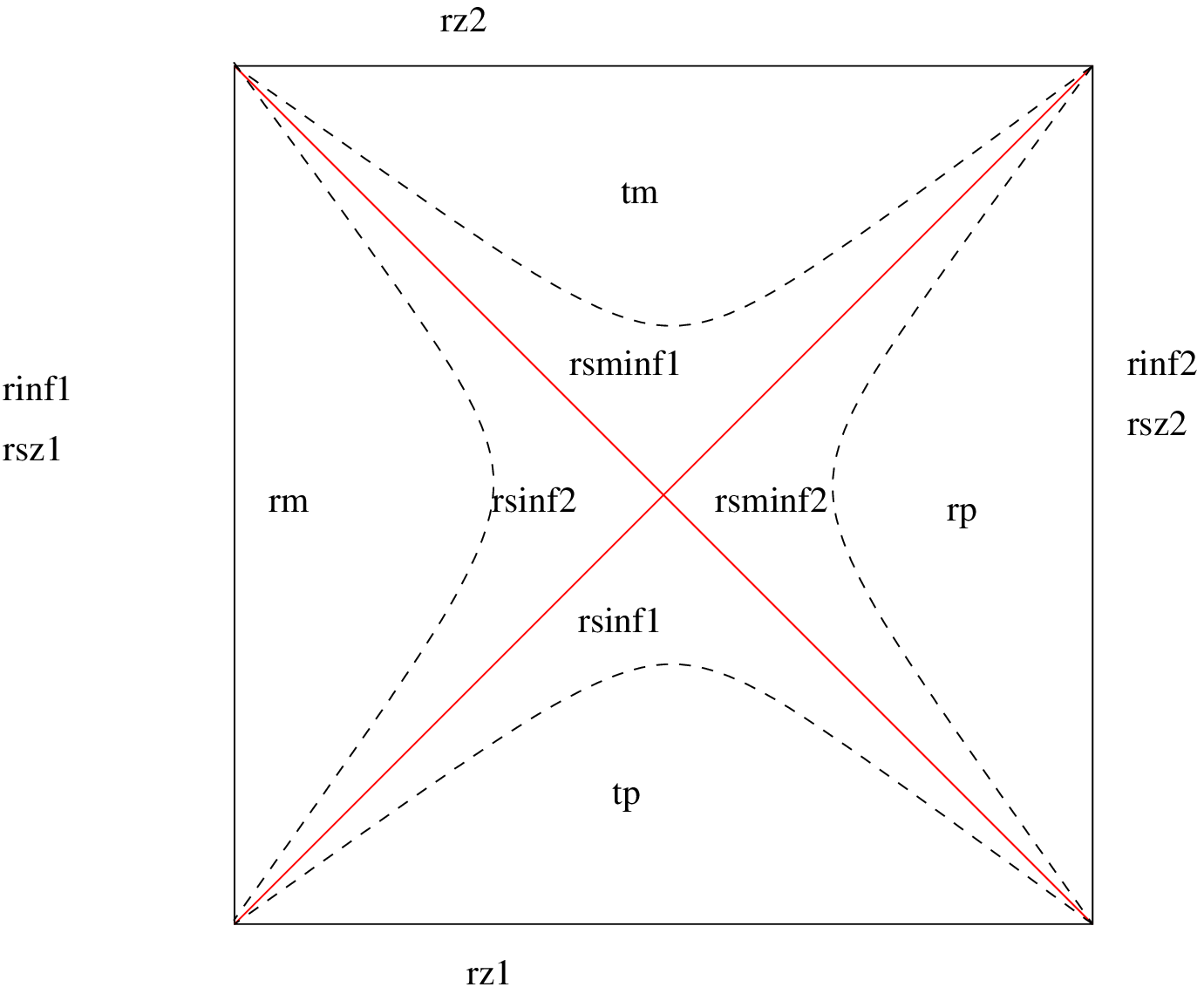}
\end{center}
\caption{}
\end{figure}
This diagram is the same as for the Schwarzschild black hole with
the only difference that instead of null infinities we have now the
time-like infinities on both sides of the Einstein-Rosen bridge. But
this black hole is strange: it has zero mass (!), and the horizon
radius $R = R_0$ is connected to the negative (!) cosmological
constant. Besides, there are no singularities at $R = 0$. The
complete Carter-Penrose diagrams for the "heavy" shell with $S^0_0
> 0$ and $S^0_0 < 0$ look as in Figs.31 and 32.

\begin{figure}[H]
\unitlength=1in
\begin{center}
\psfragscanon
\psfrag{rp1}{$R_+$}
\psfrag{rm1}{$R_-$}
\psfrag{rp2}{$R_+$}
\psfrag{rm2}{$R_-$}
\psfrag{tp1}{$T_+$}
\psfrag{tm1}{$T_-$}
\psfrag{tp2}{$T_+$}
\psfrag{tm2}{$T_-$}
\psfrag{rz1}{$R=0$}
\psfrag{rz2}{$R=0$}
\psfrag{rz3}{$R=0$}
\psfrag{rz4}{$R=0$}
\psfrag{rinf1}{$R=\infty$}
\psfrag{rinf2}{$R=\infty$}
\psfrag{rinf3}{$R=\infty$}
\includegraphics[width=4.5in]{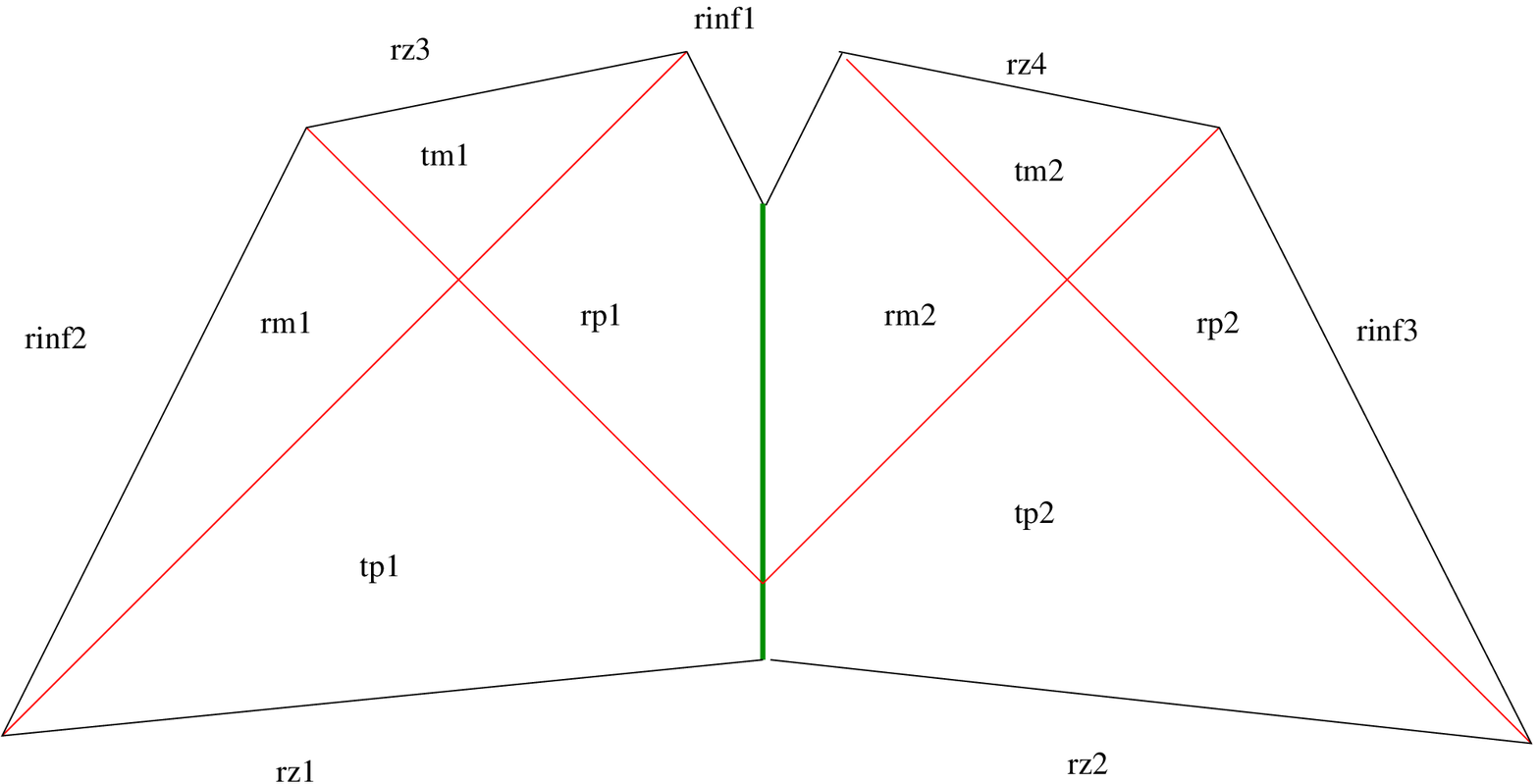}
\end{center}
\caption{}
\end{figure}

\begin{figure}[H]
\unitlength=1in
\begin{center}
\psfragscanon
\psfrag{rp}{$R_+$}
\psfrag{rm}{$R_-$}
\psfrag{tp1}{$T_+$}
\psfrag{tp2}{$T_+$}
\psfrag{r01}{$R=R_0$}
\psfrag{r02}{$R=R_0$}
\psfrag{rz}{$R=0$}
\psfrag{rinf1}{$R=\infty$}
\psfrag{rinf2}{$R=\infty$}
\psfrag{rinf3}{$R=\infty$}
\psfrag{rinf4}{$R=\infty$}
\includegraphics[width=3.5in]{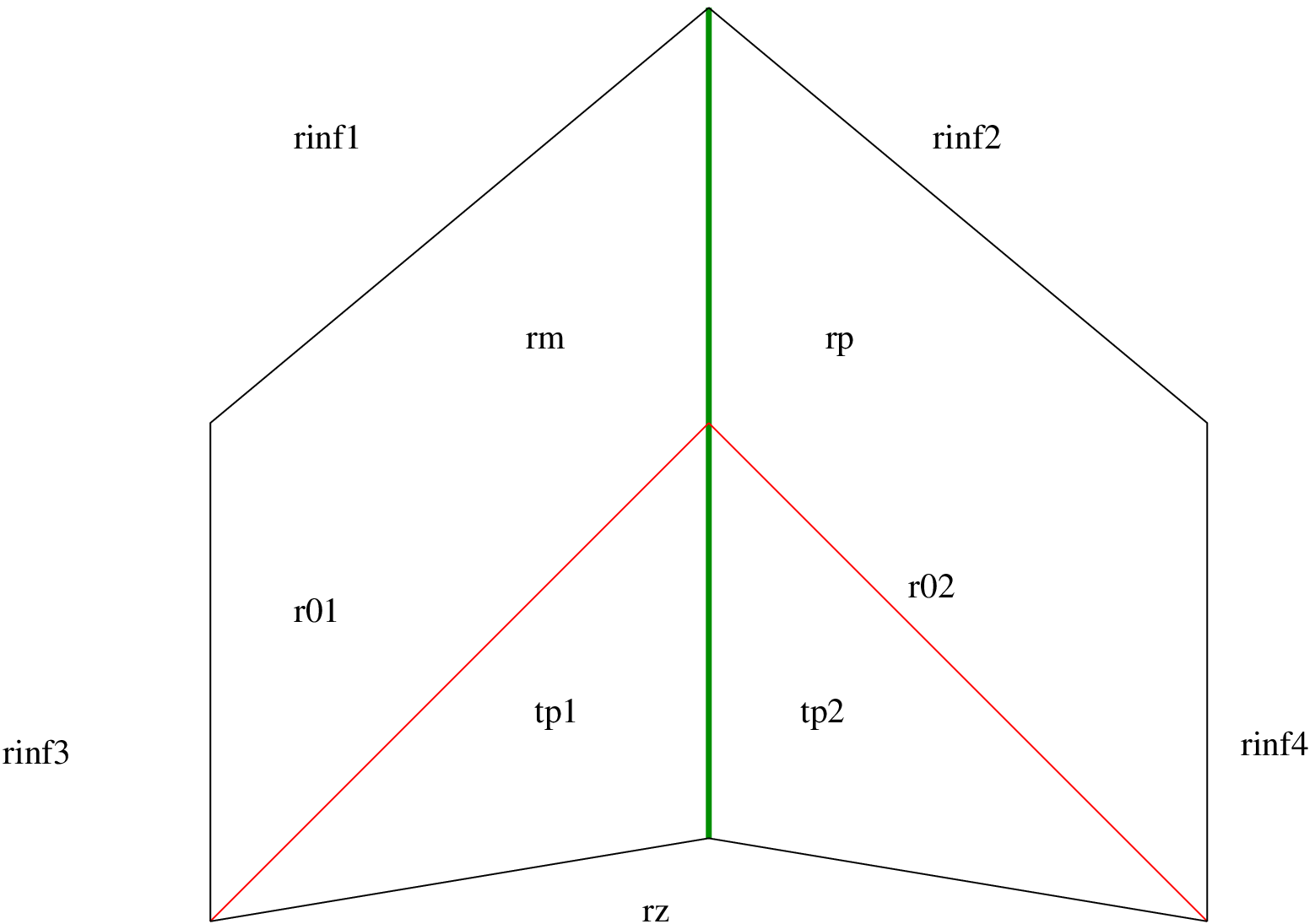}
\end{center}
\caption{}
\end{figure}
We see that in the case $S^0_0 > 0$ there zero mass black holes on
both sides of the shell and, consequently, two Einstein-Rosen
bridges. The latter property allows to have asymmetric hierarchy
without destroying the $Z_2$-symmetry of the matching. For the
"light" shell the complete Carter-Penrose diagrams are Figs.33 and
34.

\begin{figure}[H]
\unitlength=1in
\begin{center}
\psfragscanon
\psfrag{rp1}{$R_+$}
\psfrag{rm1}{$R_-$}
\psfrag{rp2}{$R_+$}
\psfrag{rm2}{$R_-$}
\psfrag{tp1}{$T_+$}
\psfrag{tm1}{$T_-$}
\psfrag{tp2}{$T_+$}
\psfrag{tm2}{$T_-$}
\psfrag{rz1}{$R=0$}
\psfrag{rz2}{$R=0$}
\psfrag{rinf1}{$R=\infty$}
\psfrag{rinf2}{$R=\infty$}
\includegraphics[width=4.5in]{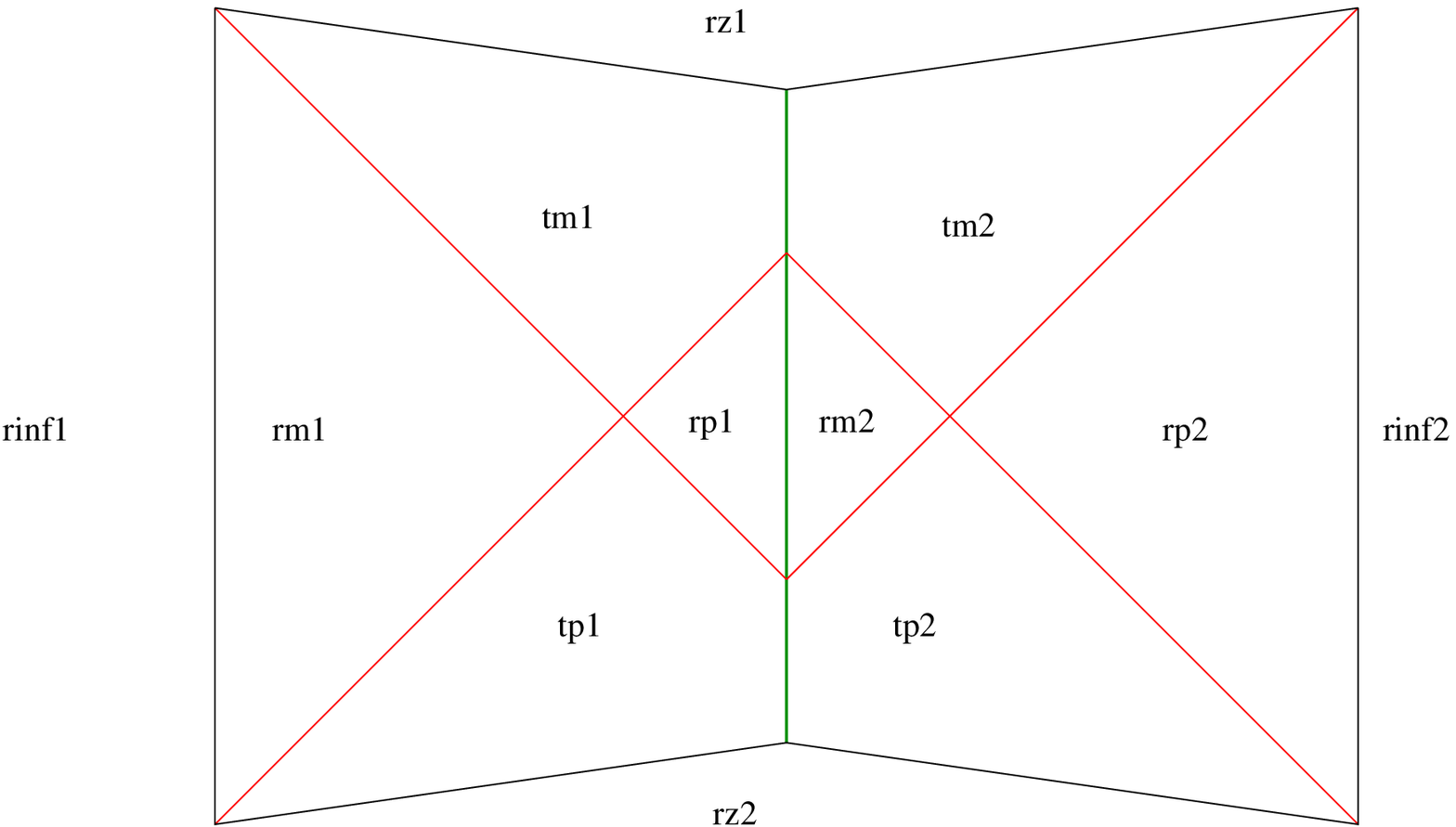}
\end{center}
\caption{}
\end{figure}

\begin{figure}[H]
\unitlength=1in
\begin{center}
\psfragscanon
\psfrag{rp}{$R_+$}
\psfrag{rm}{$R_-$}
\psfrag{tp1}{$T_+$}
\psfrag{tm1}{$T_-$}
\psfrag{tp2}{$T_+$}
\psfrag{tm2}{$T_-$}
\psfrag{rz1}{$R=0$}
\psfrag{rz2}{$R=0$}
\psfrag{rinf1}{$R=\infty$}
\psfrag{rinf2}{$R=\infty$}
\includegraphics[width=3.5in]{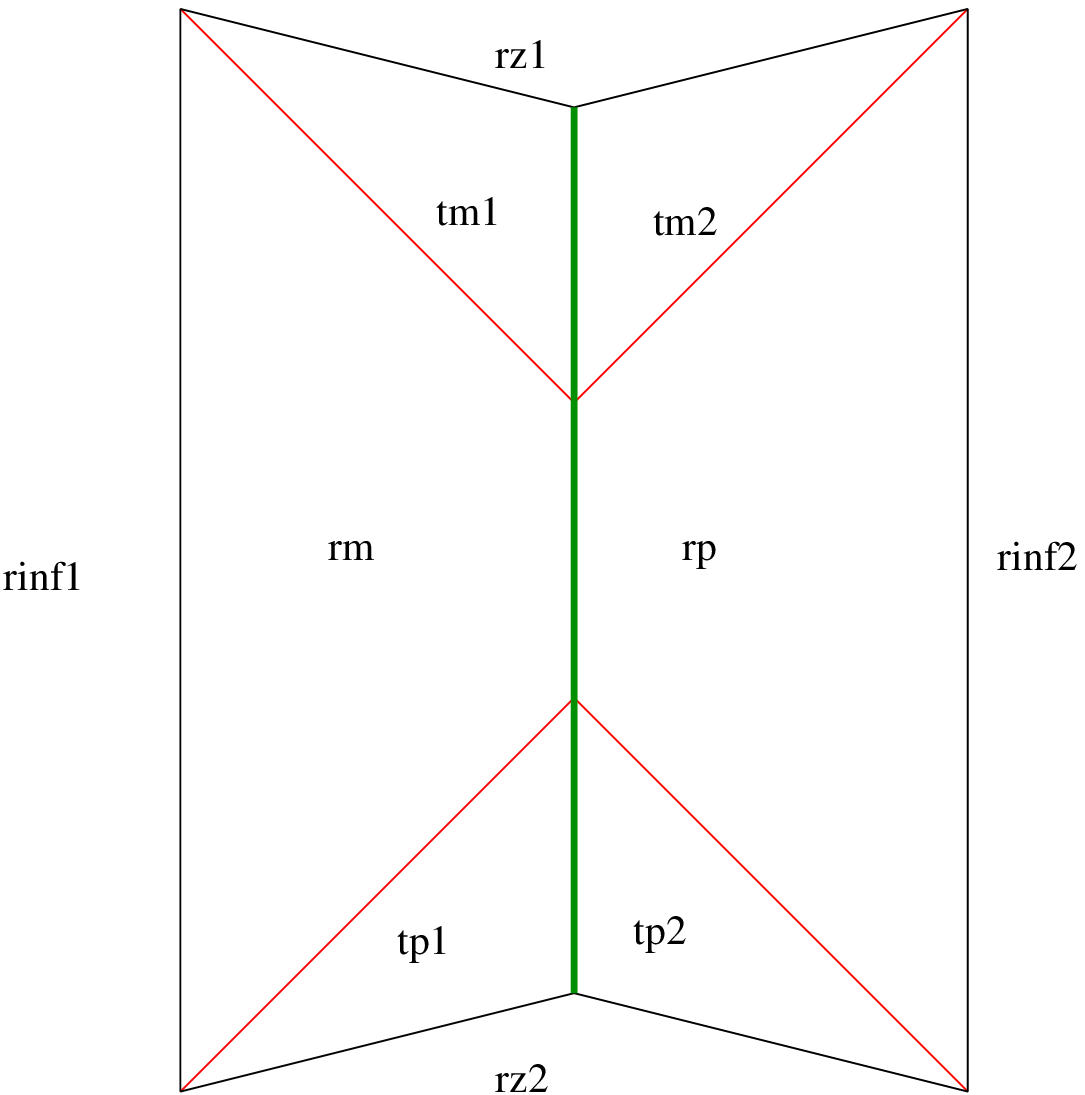}
\end{center}
\caption{}
\end{figure}
For both signs of $S^0_0$ our shell undergoes a bound motion and,
thus, can be quantized in the same way as the bound states, say, of
hydrogen atom. And the last note: because the proper time is finite
when traveling from an initial state at $R = 0$ to the final state
at $R = 0$, we can use the unfolded description in the time
direction (both past and future) as well.

We would like now to summarize what we learned studying the global
geometry of the brane universe cosmological models. About the
assumption. The most severe one is that about the existence of
cosmological symmetry throughout the whole space-time. This means
that the brane does not affect the local bulk geometry, in other
words, the latter does not depend on the place of brane matching (in
terms of invariant radius $R$). Thus, the singular shell does not
send any (gravitational) signal about its existence. The very
interesting result is the connection between the spatial curvature
of a homogeneous space on the brane and the global geometry. This is
rather unexpected and contradicts, in a sense, our four-dimensional
experience and intuition. We are used ti think  of the thin shells
as the two-dimensional bubble walls embedded into a
three-dimensional space. Because these walls are spherically
symmetric (assuming "cosmological" symmetry) we get automatically $k
= 1$. It is appeared unexpected also the possibility of
non-symmetric (from the global geometry point of view) inclusion
into consideration of several additional branes without destroying
the $Z_2$-symmetric matching of "our" brane to the bulk. This may be
useful in attempts to understand the observed non-symmetric
hierarchy of the fundamental interactions. We saw that such a
property exists (even for zero Schwarzschild mass) if we adopt some
unfolded description of the space-times with negative cosmological
constant, and also in the case $\Lambda < 0 \, k = - 1$, where we
found the Einstein-Rosen bridge which exactly like in the
Schwarzschild black hole space-time, but with the cosmological
constant playing the role of the nonzero mass. This last case is
also interesting for constructing more realistic models because it
allows a transition from the "light" shells (with the surface energy
density bounded from above) to the "heavy" shells (surface energy
density bounded from below). Besides, the quantized "light" shells
will form the bound states and this may become very important in
investigating the quantum models with several branes which could
exhibit the hierarchy features.


\begin{thebibliography}{10}
\bibitem{c1} V.A.Rubakov, M.Yu.Shaposhnikov. Phys.Lett. B125 (1083) 136-138
\bibitem{c2} L.Randall, R.Sundrum. Phys.Rev.Lett.83 (1999) 4690-4693
\bibitem{c3} L.Randall, R.Sundrum. Phys.Rev.Lett.83 (1999) 3370-3373
\bibitem{c4} I.D.Novikov. Soobsheniya GAISH 132 (1964) 43
\bibitem{c5} V.A.Berezin,  V.A.Kuzmin, I.I.Tkachev. Phys.Lett. B120 (1983) 91-96
\bibitem{c6} V.A.Berezin,  V.A.Kuzmin, I.I.Tkachev. Phys.Rev. D36 (1987) 2919-2944
\end{thebibliography}
\end{document}